\documentclass[final,1p,times]{elsarticle}

\usepackage{amssymb}
\usepackage{amsmath}
\usepackage{units}
\usepackage{color}
\usepackage{epstopdf}
\usepackage{url}

\biboptions{authoryear,sort&compress}

\usepackage{graphicx}
\newcommand{\boldface}[1]{\mathbf{#1}}

\newcommand{\bfr}{\boldface{r}}

\newcommand{\beps}{\boldsymbol{\varepsilon}}

\newcommand{\bfmu}{\boldsymbol{\mu}}

\newcommand{\bfsigma}{\boldsymbol{\sigma}}

\newcommand{\bepsp}{\boldsymbol{\varepsilon}_\mathrm{p}}

\newcommand{\dbepsp}{\dot{\boldsymbol{\varepsilon}}_\mathrm{p}}
\newcommand{\sigy}{\sigma_\mathrm{y}}
\newcommand{\sigyi}{\sigma_{\mathrm{y}i}}

\newcommand{\be}{\begin{equation}}
\newcommand{\ee}{\end{equation}}
\newcommand{\bea}{\begin{eqnarray}}
\newcommand{\eea}{\end{eqnarray}}
\newcommand{\bes}{\begin{equation*}}
\newcommand{\ees}{\end{equation*}}
\newcommand{\beas}{\begin{eqnarray*}}
\newcommand{\eeas}{\end{eqnarray*}}
\newcommand{\D}{\displaystyle}

\newcommand{\bfm}[1]{ \mathbf{#1} }
\def\dd{\;\!\mathrm{d}}
\def\dev{\mathrm{dev}}
\def\tr{\mathrm{tr}}
\def\sym{\mathrm{sym}}
\renewcommand{\inf}[3]{
\underset{#1}{\mathrm{inf}}\left\lbrace #2 \; \left| \vphantom{#2}\vphantom{#3} \right. \; #3 \right\rbrace
}
\newcommand{\infr}[2]{
\underset{#1}{\mathrm{inf}}\left\lbrace #2 \right\rbrace
}

\begin{document}

\begin{frontmatter}



\title{A variational approach to effective models for inelastic systems}


\author[a]{Ghina Jezdan}
\ead{Ghina.Jezdan@rub.de}
\author[b]{Sanjay Govindjee}
\ead{s\_g@berkeley.edu}
\author[a]{Klaus Hackl}
\ead{Klaus.Hackl@rub.de}
\address[a]{Institute of Mechanics of Materials, Ruhr-Universit\"{a}t Bochum, D-44780
	Bochum, Germany
}
\address[b]{Structural Engineering, Mechanics, and Materials; Department of Civil and Environmental Engineering, University of California, Berkeley, Berkeley CA 94720-1710, USA
}

\begin{abstract}
		
	Given a set of inelastic material models, a microstructure, a macroscopic structural geometry, and a set of boundary conditions, one can in principle always solve the governing equations to determine the system's mechanical response.  However, for large systems this procedure can quickly become computationally overwhelming, especially in three-dimensions when the microstructure is locally complex.  In such settings multi-scale modeling offers a route to a more efficient model by holding out the promise of a framework with fewer degrees of freedom, which at the same time faithfully represents, up to a certain scale, the behavior of the system.  
	In this paper, we present a methodology that produces such models for inelastic systems upon the basis of a variational scheme.  The essence of the scheme is the construction of a variational statement for the free energy as well as the dissipation potential for a coarse scale model in terms of the free energy and dissipation functions of the fine scale model.  From the coarse scale energy and dissipation we can then generate coarse scale material models that are computationally far more efficient than either directly solving the fine scale model or by resorting to FE$^2$ type modeling. Moreover, the coarse scale model preserves the essential mathematical structure of the fine scale model.  
	An essential feature for such schemes is the proper definition of the coarse scale inelastic variables.  By way of concrete examples, we illustrate the needed steps to generate successful models via application to problems in classical plasticity, included are comparisons to direct numerical simulations of the microstructure to illustrate the accuracy of the proposed methodology.
	
\end{abstract}


\end{frontmatter}


%
\section{Introduction}
\label{intro}

Engineering systems often possess multiscale character. Examples are structures with internal mechanisms or elaborate design, but also complex materials with microstructure. To model these systems to their full extent is often not desirable considering the large computation times and problems of data storage involved, but also difficulties in interpreting the obtained results. Therefore, it is of high interest to devise effective models capturing the essential behavior as close as possible while relying only on a small set of essential variables.

Focusing on microstructured materials, there are two common analysis methods seen in the literature, ones based on phenomenological models and ones incorporating direct numerical simulation (DNS).   Phenomenological models attempt to mimic the material behaviour directly at the {macroscale}. While this approach leads to models which are computationally quite efficient, the observed highly complex behaviour can require the usage of a {large number of parameters}. Often these parameters do not have clear physical meaning and are difficult to identify. This makes it almost {impossible to use these models in a predictive way} because parameters identified in a specific experiment might not be suitable in a different setting.
{DNS methods} on the other hand rely on {physically accurate theories at the microscale} using only a few parameters which are mechanistically well defined.   While these models work efficiently at the microscale, it is possible to generate information at the macroscale only via computationally expensive procedures. The effort as well as the amount of data to be stored can also be so large that it becomes practically {impossible to model any sort of macroscopic system} in a meaningful way. Moreover, a huge quantity of information is generated at the microscale, and most of it is of no deep relevance.

Both approaches are seen to be deficient for different reasons: either the models are efficient, but not reliable, or they are well rooted in physics and accurate, but computationally too costly. Both of them tend to give results which are difficult to interpret. 
In this paper we present a variational approach applied to the homogenization of inelastic materials with microstructure, that is suitable to overcome the mentioned issues.  The approach rests on the construction of both an upscaled energy density and an upscaled dissipation potential.  The resulting effective models possess the following features:
\begin{itemize}
	\item they involve only parameters which are physically well-defined and measurable,
	\item the state-equations can be evaluated without resorting to elaborate computational schemes,
	\item essential material behaviour at the microscale as well as the macroscale is captured.
\end{itemize}
The general notion of variational homogenization for inelastic systems is not entirely new.
Models with similar properties have been derived for example for upscaling martensitic phase transformations in shape memory alloys \citep{govindjee2007upper,hackl2008micromechanical,junker2011finite,junker2014thermo,WAIMANN2016110} and  in the prediction of dislocation patterns in single crystal metals \citep{kochmann2011evolution,hackl2012zamm,gunther2015variational,Aubry20032823,ContiDolzmannKlust2009,ContiOrtiz05,miehe2010,
	miehe2011a,miehe2003,miehe2004}. More recently, an application to finite-deformation viscoelasticity has appeared in \cite{govindjee2019fully}, wherein the concept of an upscaled dissipation potential also appears within the restricted confines of the microsphere model \citep{miehe.ea:04}.

When applied to the homogenization of periodic elastoplastic microstructures, a close relation between our proposed approach and the so-called
transformation field analysis {(TFA)} can be observed. TFA is a computational homogenization method established in \cite{Dvorak1992,DvorakBenveniste1992}. Similar to the present proposal, the overall domain in TFA is divided into subdomains. However, differently than here, pre-defined piece-wise linear inelastic local strain fields are assumed.  Further in TFA and its more modern variant, nonuniform TFA (NTFA), the inelastic fields must be resolved at the microscale using microscopic evolution equations \citep{MichelSuquet2003,FRITZEN2015114}.  In our proposal, we provide a means for the analytic generation of the evolution equations for macroscopic inelastic variables, a step that helps to reduce data storage demands.

\section{Variational framework}\label{sec:vf}

We introduce a variational approach for the description of inelastic processes resting on thermodynamical extremum principles. For this purpose let us consider a physical system described by (sets of) external, i.e.\ controllable, state variables $\bfm{x}$ and internal state variables $\bfm{z}$, parametrized as
\be
\label{eq1}
{\bfm{x}}={\bfm{x}}(\xi), \quad {\bfm{z}}={\bfm{z}}(\xi), \qquad \xi\in\Omega,
\ee
where $\Omega$ denotes a suitable parameter space, e.g. physical space, a set  of directions, or a probability space.

We will assume that the system's behaviour may be defined using only two scalar potentials: a free energy $\Psi(\nabla\bfm{x},\bfm{x},\bfm{z})$ and a dissipation potential $\Delta(\bfm{z},\dot{\bfm{z}})$, where $\nabla$ denotes the gradient with respect to $\xi$.

The external state variable $\bfm{x}$ will be given by minimization of the system's potential energy as in
\be
\label{eq2}
\inf{\bfm{x}}{\int_\Omega \Psi(\nabla\bfm{x},\bfm{x},\bfm{z}) \dd\xi + f_\mathrm{ext}(\bfm{x})}{\bfm{x}=\bfm{x}_0 \;\mbox{on}\; \partial\Omega},
\ee
where $f_\mathrm{ext}(\bfm{x})$ denotes the potential of external driving forces. The evolution of the internal variables is described by the Biot-equation
\be
\label{eq3}
\frac{\partial \Psi}{\partial {\bfm{z}}}+\frac{\partial \Delta}{\partial \dot{\bfm{z}}} = {\bf 0}.
\ee
Note that Eq.\  \eqref{eq3} is the stationarity condition of the minimization problem
\be
\label{eq4}
\infr{\dot{\bfm{z}}}{\dot{\Psi} + \Delta}.
\ee

Our goal is to capture the behavior of the system described by Eqs.\  \eqref{eq2} and \eqref{eq4} using only a finite (small) number of parameters. Assume that our state variables are, as functions of $\xi$, members of suitable function spaces: $\bfm{x}\in X$, $\bfm{z}\in Z$. Let us moreover introduce \emph{linear projection operators} to finite-dimensional spaces
\be
\label{eq5}
\mathrm{P}: X \longrightarrow \mathbb{R}^M, \qquad \mathrm{Q}: Z \longrightarrow \mathbb{R}^N.
\ee
These projection operators define \emph{essential parameters} via
\be
\label{eq6}
\bfm{x}_\mathrm{ess}=\mathrm{P}\bfm{x}, \qquad \bfm{z}_\mathrm{ess}=\mathrm{Q}\bfm{z},
\ee
and spaces of \emph{marginal remainders} as kernels of the operators:
\be
\label{eq7}
X_\mathrm{mar} = \left\lbrace \bfm{x} \in X \,|\, \mathrm{P}\bfm{x} = \mathbf{0} \right\rbrace, \qquad 
Z_\mathrm{mar} = \left\lbrace \bfm{z} \in Z \,|\, \mathrm{Q}\bfm{z} = \mathbf{0} \right\rbrace.
\ee
We would like the kinetics of the system under consideration to be captured by the essential parameters as closely as possible. This occurs when the potentials $\Psi$ and $\Delta$ are invariant under variations within the marginal spaces. Hence, the ideal situation arises when
\be
\label{eq8}\begin{aligned}
&\frac{\partial \Psi}{\partial \bfm{x}}:\delta\bfm{x} = 0 \; \mbox{for} \; \delta\bfm{x} \in X_\mathrm{mar}, \qquad
&&\frac{\partial \Psi}{\partial \bfm{z}}:\delta\bfm{z} = 0 \; \mbox{for} \; \delta\bfm{z} \in Z_\mathrm{mar}, \qquad\\
&\frac{\partial \Delta}{\partial \bfm{z}}:\delta\bfm{z} = 0 \; \mbox{for} \; \delta\bfm{z} \in Z_\mathrm{mar}, \qquad
&&\frac{\partial \Delta}{\partial \dot{\bfm{z}}}:\delta\dot{\bfm{z}} = 0 \; \mbox{for} \; \delta\dot{\bfm{z}} \in Z_\mathrm{mar},\end{aligned}
\ee
where a colon denotes a suitable inner product. In a general situation this does not happen.  However, Eqs.\ \eqref{eq8} are nothing else than the stationarity conditions of specific minimization problems. Thus, we use  Eqs.\ \eqref{eq8} to motivate macroscopic versions of the potentials, ones that are in fact invariant to variations in the marginal spaces:
\begin{align}
\label{eq9}
\Psi_\mathrm{macro}(\bfm{x}_\mathrm{ess},\bfm{z}_\mathrm{ess}) &=
\inf{\bfm{x} \in X, \bfm{z} \in Z}{\frac{1}{|\Omega|}\int_\Omega \Psi(\nabla\bfm{x},\bfm{x},\bfm{z}) \dd\xi}{ \bfm{x}_\mathrm{ess}=\mathrm{P}\bfm{x}, \bfm{z}_\mathrm{ess}=\mathrm{Q}\bfm{z}}, \\
\label{eq10}
\Delta_\mathrm{macro}(\bfm{z}_\mathrm{ess},\dot{\bfm{z}}_\mathrm{ess}) &=
\inf{\bfm{z} \in Z, \dot{\bfm{z}} \in Z}{\frac{1}{|\Omega|}\int_\Omega \Delta(\bfm{z},\dot{\bfm{z}}) \dd\xi}{ \bfm{z}_\mathrm{ess}=\mathrm{Q}\bfm{z}, \dot{\bfm{z}}_\mathrm{ess}=\mathrm{Q}\dot{\bfm{z}}}.
\end{align}
As a further key constructive \emph{assumption}, the minimization problem in Eq.\  \eqref{eq4} is replaced by
\be
\label{eq11}
\infr{\dot{\bfm{z}}_\mathrm{ess} \in \mathbb{R}^N}{\dot{\Psi}_\mathrm{macro} +  \Delta_\mathrm{macro}}.
\ee
Lastly, we assume that the potential of external loads in Eq.\  \eqref{eq2} can be expressed via the essential parameters, $f_\mathrm{ext}(\bfm{x}) = f_\mathrm{ess}(\bfm{x}_\mathrm{ess})$, and that the boundary conditions may be expressed in terms of the essential parameters, too, by specifying a projection $\mathrm{B}\bfm{x}_\mathrm{ess}$.   In this setting, we replace the minimization problem in Eq.\  \eqref{eq2} with the macroscopic form:
\be
\label{eq12}
\inf{\bfm{x}_\mathrm{ess} \in \mathbb{R}^M}{\int_\Omega \Psi_\mathrm{macro}(\bfm{x}_\mathrm{ess},\bfm{z}_\mathrm{ess}) \dd\xi + f_\mathrm{ess}(\bfm{x}_\mathrm{ess})}{\mathrm{B}\bfm{x}_\mathrm{ess}=\bfm{x}_{\mathrm{ess}0} \;\mbox{on}\; \partial\Omega}.
\ee
In what follows, the two minimization problems in Eqs.\  \eqref{eq11} and \eqref{eq12} will be shown to capture the kinetics  of the macroscopic system, thus approximating the behavior of the original system. Given $f_\mathrm{ess}$ and $\bfm{x}_{\mathrm{ess}0}$ as functions of time, they allow for the computation of $\bfm{x}_\mathrm{ess}$ and $\bfm{z}_\mathrm{ess}$ as functions of time.

\section{The variational formulation of elastoplasticity with isotropic hardening}\label{sec:3}

We will demonstrate the theory introduced above by applying it to bodies having standard small-strain rate-independent elastoplastic material behavior. In this case $\Omega \subset \mathbb{R}^n$ is just a domain in Euclidean space representing a material body. The external variable $\bfm{x} = \bfm{u}=\bfm{u}(\bfm{r},t)$ is the displacement field as a function of the position vector $\bfm{r} \in \Omega$ and the time parameter $t$. The internal variables are given by $\bfm{z}=\{\bepsp,q\}$, where $\bepsp$ with $\tr \,\bepsp=0$ is the plastic strain and $q$ is a scalar hardening variable. The free energy is given as
\be
\label{eq13}
\Psi(\beps,\bepsp) = \frac{1}{2} \left( \beps - \bepsp \right) : \mathbb{C} : \left( \beps - \bepsp \right) + \frac{1}{2} \, a \, q^2,
\ee
where $\beps = \frac{1}{2} \left( \nabla\bfm{u} + \nabla\bfm{u}^\mathrm{T} \right)$ is the total strain,  $\mathbb{C}=\mathbb{C}(\bfm{r})$ is the tensor of elastic moduli and $a(\bfm{r})$ is the hardening modulus. The dissipation potential has the form
\be
\label{eq14}
\Delta(\dbepsp) = \sqrt{\frac{2}{3}} \sigy \left\| \dbepsp \right\|,
\ee
where $\sigy=\sigy(\bfr)$ is the initial yield stress. The evolution of $q$ is governed by the constraint
\be
\label{eq14-1}
\dot q(\bfm{r},t) = \left\| \dbepsp \right\|.
\ee
Employing constraint Eq.\ \eqref{eq14-1}, we can construct a Lagrangian to use in the extremal problem Eq.\ \eqref{eq4}:
\be
\label{eq14-2}
L = \dot{\Psi} + \Delta + \gamma\left(\dot{q} - \left\| \dbepsp \right\|\right)= \bfsigma : \left( \dot \beps - \dbepsp \right) + a \, q \dot{q} + \sqrt{\frac{2}{3}} \sigy \left\| \dbepsp \right\| + \gamma\left(\dot{q} - \left\| \dbepsp \right\|\right),
\ee
where $\bfsigma = \mathbb{C} : \left( \beps - \bepsp \right)$ is the stress tensor. Variations with respect $\gamma$, $\dot{q}$, and $\dbepsp$ yield the Biot-equation, Eq.\ \eqref{eq3}: 
\begin{align}
\dot q(\bfm{r},t) &= \left\| \dbepsp \right\|  \\
a\, q + \gamma &= 0\\
\label{eq14-3}
\dev \bfsigma &\in \left( \sqrt{\frac{2}{3}} \sigy + a \, q \right) \partial\left\| \dbepsp \right\|.
\end{align}
Note that we obtain a differential inclusion because of the non-differentiability of the norm \citep{rockafellar1970}.  Thus Eq.\ \eqref{eq14-3} includes the yield condition
\be
\label{eq14-4}
\left\| \dev \bfsigma \right\| \leq \sqrt{\frac{2}{3}} \sigy + a \, q
\ee
as well as the flow rule
\be
\label{eq14-5}
\dbepsp = \lambda \, \dev \bfsigma, \quad \lambda \geq 0.
\ee

\section{Elastoplastic structures with high symmetry}

As a first illustration of the proposed homogenization scheme we consider the problem of inelastic homogenization in structural elements.  In particular, consider a homogeneous three-dimensional elastoplastic body and the question of the determination of an elastoplastic model at the structural level. Our first example is the torsion of a right-circular cylinder and the second is the pressurization of a spherical vessel.  In both cases the (pointwise) elastoplastic models  consist of state variables in infinite dimensional function spaces.  Using the proposed construction of Sec.~\ref{sec:vf}, we show how to construct low dimensional approximations for the structural level (macroscopic) behavior.

\subsection{Cylindrical body subject to torsion: boundary value problem}
\label{cyl}

\begin{figure}
\centering
  \includegraphics[scale=0.55]{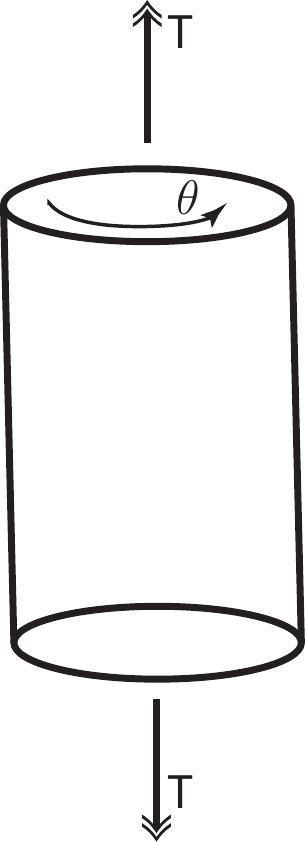}  
    \caption{Torsion bar problem, a cylinder with radius $R$ subject to twist $\theta$ or torque $T$.}
    \label{fig0a}
 \end{figure}

Assume $\Omega$ to be a cylindrical body with radius $R$, subject to either end torques $T$ or an end rotation $\theta$ (see Fig.~\ref{fig0a}), which implies a classical displacement field with strictly rotational motion, 
\be
\label{eq79}
u=u_\phi(r,z,t) = r \, z \, \theta^\prime(t)\,,
\ee 
where $t$ is a time parameter, $z$ is the axial coordinate, $r$ is the radial coordinate, and $\theta^\prime(t) := \theta(t)/L$ is the twist, i.e.\ the rotation angle per unit length of the cylinder with length $L$.  In what follows, we will generally suppress the time argument.
The response of such a cylinder to imposed torque or twist for a perfectly elastoplastic material is a well understood problem, and is commonly introduced in elementary mechanics texts; see e.g.\ \citet[][Sec.~7.6]{govindjee2013}.  Missing from such presentations, is an explicit low dimensional model relating the cylinder's torque (a scalar) to its twist (a scalar) and a minimal number of internal variables (scalars) characterizing the inelastic response such that arbitrary loadings in time may be effectively modeled.  Using the framework outlined above, it is possible to achieve such a model.

Applying cylindrical coordinates, the non-trivial strain components are:
\be
\label{eq80}
\gamma := 2\varepsilon_{z\phi} =2\varepsilon_{\phi z}= \theta^\prime r ,
\ee
where $\gamma$ is the non-trivial engineering shear strain in the cylinder. In addition, we introduce the plastic strain component $ p=p(r,t)$,  
\be
\label{eq81}
p:=2\varepsilon_{\mathrm{p}z\phi} =2 \varepsilon_{\mathrm{p}\phi z}.
\ee
The shear stress field in the cylinder is
\be
\tau(r):= \sigma_{z\phi}=\sigma_{\phi z} = \mu(\gamma - p),
\ee
where $\mu$ is the shear modulus of the material. The free energy and dissipation potential in Eqs.\ \eqref{eq13} and \eqref{eq14} reduce to
\be
\Psi = \frac{1}{2} \mu (\gamma-p)^2 + \frac{1}{2} \; a q^2, \quad \Delta(\dot p) = \tau_y  \left| \dot p \right|.
\ee
where $\tau_y$ is the yield stress in shear. 
Yield condition and flow rule given in Eqs.\ \eqref{eq14-4} and \eqref{eq14-5} become
\be
\left| \tau \right| \le \tau_y + a \, q, \quad \dot p = \lambda \, \tau, \quad \lambda \ge 0.
\ee
The evolution of hardening in Eq.\ \eqref{eq14-1} reduces to
\be
\label{eq81a}
\dot q(r,t) = \left| \dot p(r,t)  \right|.
\ee
The closure of the system of equations, involves the specification of an imposed end rotation $\theta(t)$, equivalently $\theta'(t)$, or an
imposed end-torque
\be
T = \int_0^R 2\pi \tau r^2\, dr,
\ee
which is treated in the sense of St.\ Venant.  Given either boundary condition, the boundary value problem involves solving for  either $\theta'(t)$ or $T(t)$, as well as $p(r,t)$, $\gamma(r,t)$, and $\tau(r,t)$ -- observing that $u=rz\theta^\prime$.  Note that the primary internal variables, $p(r,t)$ and  $q(r,t)$, are functions and are thus members of an infinite dimensional space.

\subsubsection{Effective torsion model}

In order to construct a low dimensional version of the model outlined in the previous section, 
we consider subdividing $\Omega$ into $N_\mathrm{sd}$ sub-domains over the radius of the cylinder; in other words,
the cylinder is split into concentric rings:
\be
\label{eq82}
\Omega= \bigcup_{i=1}^{N_\mathrm{sd}} \Omega_{i} .
\ee
Let us for convenience introduce the moments
\be
\label{eq82a}
m^{(1)}_i = \int_{r_{i-1}}^{r_i} r \,\dd r = \frac{(r_i^2-r_{i-1}^2)}{2}, \quad m^{(2)}_i = \int_{r_{i-1}}^{r_i} r^2 \,\dd r = \frac{(r_i^3-r_{i-1}^3)}{3}.
\ee
Averages at each sub-domain can be computed using the equation
\be
\label{eq83}
\langle f \rangle_i =  \frac{1}{|\Omega_i|} \int_{\Omega_i} f \dd V= {\frac{1}{m^{(1)}_i}} \, \int_{r_{i-1}}^{r_i} f r \dd r\,,
\ee
where
\be
\label{eq84}
r_\mathrm{0}=0< r_1< \ldots < r_{N_\mathrm{sd}-1} < r_{N_\mathrm{sd}}=R.
\ee
The mean values for the plastic parameter and linear hardening per each sub-domain are given as
\be
\label{eq85}
p_i =  \langle p \rangle_i, \quad
q_i= \langle q \rangle_i\,,
\ee
and the essential internal parameters are assumed to be the linear projections
\be
\label{eq86} 
\bfm{z}_\mathrm{ess} = \{p_i, q_i\}\qquad i\in\{1,2,\ldots,N_\mathrm{sd}\}\,. 
\ee
Whereas the essential external parameter is simply
\be
\label{eq87}
\bfm{x}_\mathrm{ess} = \{\theta^\prime\}.
\ee
Due to the assumption of Eq.\ \eqref{eq79}, Eq.\ \eqref{eq87} provides an exact representation of the displacement field. Put another way, the problem is fully kinematically driven, which greatly simplifies the analysis --- the normally infinite dimensional state space for $\bfm{x}$ is, in this example, one-dimensional and exactly represented.

The macroscopic (or effective) free energy, 
Eq.\ \eqref{eq9} 
, now takes the form
\begin{multline}
\label{eq88}
 \Psi_\mathrm{macro}(\theta^\prime,p_1,\ldots,p_{N_\mathrm{sd}},q_1,\ldots,q_{N_\mathrm{sd}} )  \\
 =\inf{p,q}{\int_{0}^{R} 2\pi \left[ \frac{1}{2}\, \mu \left(r \theta^\prime- p \right)^2 +\frac{1}{2}a\;q^2\right] r \dd r} {\langle p \rangle_i =p_i,\;\langle q \rangle_i=q_i}.
\end{multline}
 Employing Lagrange-parameters $\lambda_i$, $\xi_i$ for the constraints, the Lagrangian corresponding to the minimization in Eq.\ \eqref{eq88} is given by
\begin{multline}
\label{eq88b}
L = \sum_{j=1}^{N_\mathrm{sd}}  \left[
\int_{r_{j-1}}^{r_j} 2\pi \left( \frac{1}{2}\, \mu \left(r \theta^\prime- p \right)^2 +\frac{1}{2}a\;q^2\right) r \dd r + \lambda_j \left( p_j - \frac{1}{m^{(1)}_j} \, \int_{r_{i-1}}^{r_i} p r \dd r \right) \right. \\ \left. + \xi_j \left( q_j - \frac{1}{m^{(1)}_j} \, \int_{r_{i-1}}^{r_i} q r \dd r \right) \right].
\end{multline}
The stationarity conditions with respect to $p$ and $q$ yield
\be
\label{eq88c}
p = r \theta^\prime + \frac{1}{2\pi\mu} \, \frac{\lambda_j}{m^{(1)}_j}, \quad
q = \frac{1}{2\pi a} \, \frac{\xi_j}{m^{(1)}_j} \quad \text{in} \quad \Omega_j.
\ee
Evaluation of the constraints gives
\be
\label{eq88d}
p_j = \frac{m^{(2)}_j}{m^{(1)}_j} \, \theta^\prime + \frac{1}{2\pi\mu} \, \frac{\lambda_j}{m^{(1)}_j}, \quad
q_j = \frac{1}{2\pi a} \, \frac{\xi_j}{m^{(1)}_j}.
\ee
Finally, substitution of Eqs.\ \eqref{eq88c} and \eqref{eq88d} into Eq.\ \eqref{eq88} gives
\be
\label{eq88e}
\Psi_\mathrm{macro}(\theta^\prime,p_1,\ldots,p_{N_\mathrm{sd}},q_1,\ldots,q_{N_\mathrm{sd}} ) 
=\sum_{j=1}^{N_\mathrm{sd}}  \left[
2\pi \, m^{(1)}_j \left( \frac{1}{2}\, \mu \left(\frac{m^{(2)}_j}{m^{(1)}_j} \theta^\prime- p_j \right)^2 +\frac{1}{2}a\, q_j^2\right) \right].
\ee
Let us consider now the dissipation potential:
\be
\label{eq88a}
\Delta_\mathrm{macro}(\dot p_1,\ldots,\dot p_{N_\mathrm{sd}} )  =\inf{\dot p}{\int_{0}^{R} 2\pi \; \tau_y \; \vert \dot p \vert \, r \dd r}{\langle \dot p \rangle_i=\dot p_i}.
\ee
Utilizing a Lagrange multiplier to enforce the constraints, say, $\varsigma_j$, then the stationarity conditions yield
\be
\mathrm{sign}( \dot p ) = \frac{\varsigma_j}{m_j^{(1)} 2\pi \tau_y} \qquad \text{in}~\Omega_j\,,
\ee
i.e., $\mathrm{sign}(\mathrm{\dot{p}})$ is constant in $\Omega_j$ and therefore $|\dot{p}_j| ={\vert\langle \dot p \rangle_j \vert=}\langle |\dot p| \rangle_j$;
substitution into Eq.\ \eqref{eq88a} yields
\be
\label{eq88f}
\Delta_\mathrm{macro}(\dot p_1,\ldots,\dot p_{N_\mathrm{sd}} ) 
= \sum_{j=1}^{N_\mathrm{sd}}\,2\pi \, m^{(1)}_j \tau_y \, \vert \dot p_j \vert.
\ee
Finally, observe that both minimizers, for $q(r,t)$ in Eq.\ \eqref{eq88} and for $\dot p(r,t)$ in Eq.\ \eqref{eq88a} respectively, are constant in every subdomain. Thus, Eq.\ \eqref{eq81a} simply translates into
\be
\label{eq88g}
\dot q_i = \vert \dot p_i \vert \quad\text{ for }\quad i = 1,\ldots,N_\mathrm{sd}.
\ee
We obtain the torque as
\be
\label{eq200}
T = \frac{\partial \Psi_\mathrm{macro}}{\partial \theta^\prime} = 2 \pi \mu \sum_{j=1}^{N_\mathrm{sd}}  
\left(\frac{(m^{(2)}_j)^2}{m^{(1)}_j} \theta^\prime- m^{(2)}_j p_j \right).
\ee
The effective Biot-equations as given by
\be
\label{eq201}
0 \in\frac{\partial \Psi_\mathrm{macro}}{\partial p_i}+\frac{\partial \Delta_\mathrm{macro}}{\partial \dot p_i}\,;
\ee
specifically when the constraints Eqs. \eqref{eq88g} are imposed,
\be
\label{eq202}
\mu \left(\frac{m^{(2)}_i}{m^{(1)}_i} \theta^\prime- p_i \right) \in \left( \tau_y + a q_i \right) \mathrm{sign} \, \dot p_i, \quad \text{ for } i = 1,\ldots,N_\mathrm{sd}.
\ee
The corresponding yield condition and flow rule are established as
\be
\label{eq203}
\mu \left|\frac{m^{(2)}_i}{m^{(1)}_i} \theta^\prime- p_i \right| \le \tau_y + a \, q_i, \quad \dot p_i = \lambda_i \, \left(\frac{m^{(2)}_i}{m^{(1)}_i} \theta^\prime- p_i \right), \quad \lambda_i \ge 0, \quad \text{ for } i = 1,\ldots,N_\mathrm{sd}.
\ee
 

 
 \subsubsection{Example: Torsion}
We consider an example employing Young's modulus $E= 1000 \,\mathrm{N/mm^2}$, Poison's ratio $\nu= 0.25$, yield stress $\tau_y=0.5 \sqrt{2/3} \,\mathrm{N/mm^2}$ and hardening modulus $a=0$ or $a=\mu/50$ for perfect elastoplasticity or linear hardening, respectively. The cylindrical body has the radius $R=10 \,\mathrm{mm}$ and is divided into $N_\mathrm{sd}=5$ sub-domains with equidistant radii.  
\begin{figure}
\centering
  \includegraphics[scale=0.53]{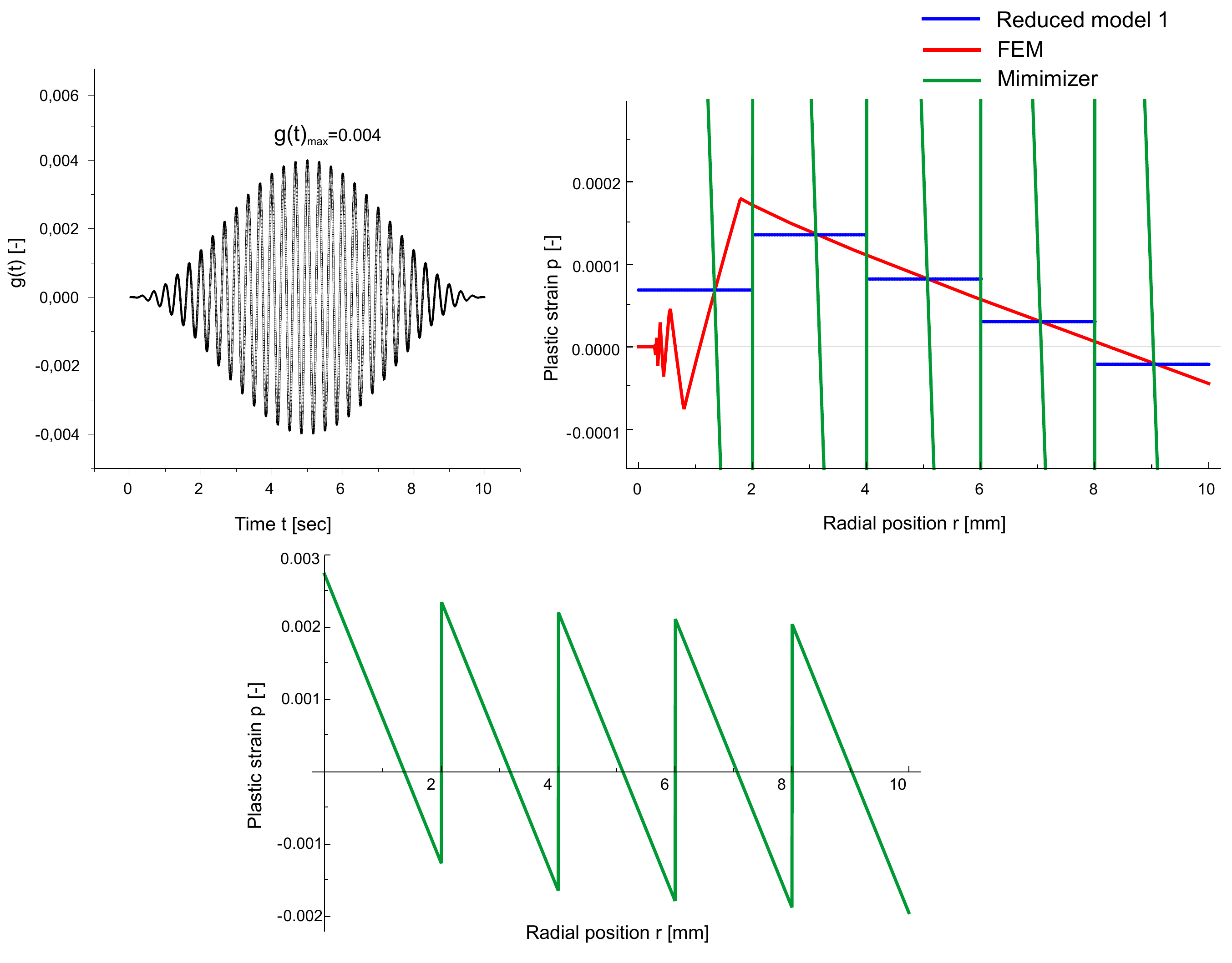}  
    \caption{Loading applied to the torsion problem $\theta^\prime = A_1 g(t)$ in the form of twist rate over time (top left) and the resulting plastic strains in the effective model versus radius $r$ at time $t= 7.5 \, \mathrm{sec}$ compared to the finite element results and the minimizer (top right) and the mimimizer's curve in a full view (bottom).}
    \label{fig0b}
 \end{figure}
We apply a cyclic loading in time for twist along the length of the cylinder $\theta^\prime = A_1 g(t)$, where $A_1= 1.0~\mathrm{rad/mm}$ and the time variation 
\be
g(t) = 2\times 10^{-3}\left (1 - \cos{\left(\frac{\pi }{5 } \frac{t}{t_0}\right)}\right) \cos{\left(\frac{30 \pi}{5} \frac{t}{t_0}\right)}
\ee
 with $t_0 = \frac{0.1\times 5}{600} \,\mathrm{sec}$, as shown in Fig.~\ref{fig0b} (top left).

The torque is calculated according to Eq.\ \eqref{eq200}. The essential internal variables $\{p_i,q_i\}$, $i = 1,\ldots,N_\mathrm{sd}$ are updated using a backward-Euler scheme employing Eq.\ \eqref{eq203}.
%


The evolution of the plastic strains in the effective model as a function of radial position $r$ at time $t= 7.5\, \mathrm{sec}$ is shown in Fig.~\ref{fig0b} (top right) in blue. Discontinuous behavior is observed as can be inferred from Eq.\ \eqref{eq88d}, in which the plastic strain is constant in each subdomain. 
\begin{figure}
\centering
  \includegraphics[width=\textwidth]{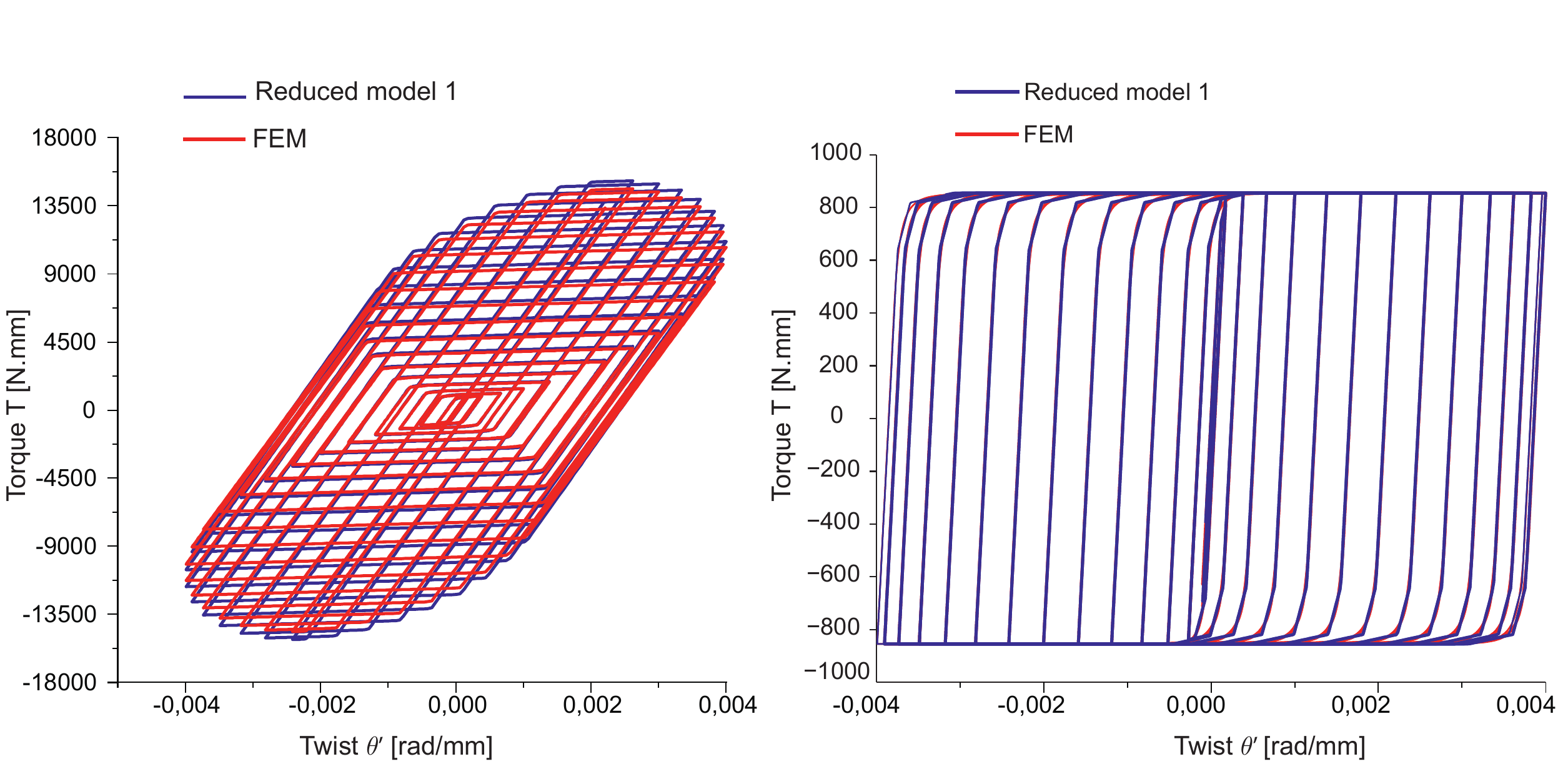}  
    \caption{Torque versus twist in the form of cyclic loading for a plastic material with isotropic hardening (left) and with perfect plasticity (right).}
    \label{fig0c}
 \end{figure}
%
%
%
The original boundary value problem has been solved using the finite element method as well, in order to generate a converged reference solution to which the results from the reduced model can be compared.  Figure \ref{fig0b} (top right) shows this solution in red.  One sees that the reduced model is providing a reasonable averaged value of the true, infinite dimensional, internal variable $p(r)$. In green in Fig.~\ref{fig0b} (bottom and top right) is the minimizing curve from Eq.~\ref{eq88c}$_1$ which is linearly dependent on the radial position and shows jumps at the interfaces of the different sub-domains. Notice that the minimizer is intersecting the curves from the reduced model exactly at the same intersection points from the finite element computations.

Torque versus twist for both perfect plasticity and plasticity with linear hardening is shown in Fig.~\ref{fig0c}. We can observe a very high level of agreement between the behavior of the effective model and the exact solution (converged FEM computation). 


%
\subsection{Spherical body under pressure}

\begin{figure}
\centering
  \includegraphics[scale=0.5]{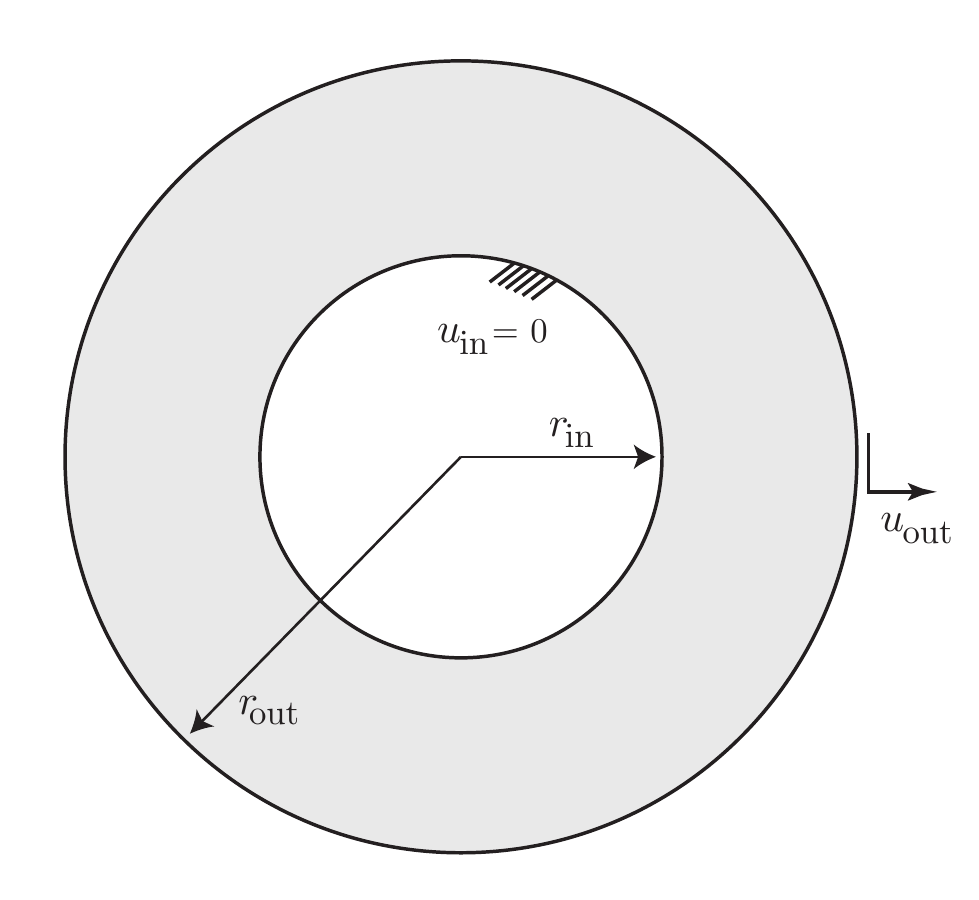}  
    \caption{Spherical body fixed at inner radius $ r_\mathrm{in}$ and displaced at outer radius $r_\mathrm{out}$.}
    \label{fig0m}
 \end{figure}
 We now consider an example in which the choice of essential external parameters is more complex. 
Let $\Omega=\{\bfm{r} \in \mathbb{R}^3 \,|\, r_\mathrm{in} \leq |\bfm{r}| \leq r_\mathrm{out}\}$ be a spherical body with inner radius $r_\mathrm{in}$ and outer radius $r_\mathrm{out}$. The sphere is pressurized internally, implying a displacement field that is dependent on the radial position $r$, $u:=u_r(r,t)$. Employing spherical coordinates, the non-trivial components of the total strain tensor are given as
\be
\label{eq100}
\varepsilon_{rr} = u^\prime, \quad \varepsilon_{\theta\theta} = \varepsilon_{\varphi\varphi} = \frac{u}{r},
\ee
where $u^\prime=\partial_r u$.
Because of the spherical symmetry of the problem, we can introduce a function $p=p(r,t)$ in order to express the non-trivial components of the plastic strain tensor as
\be
\label{eq101}
\varepsilon_{\mathrm{p}rr} = p, \quad \varepsilon_{\mathrm{p}\theta\theta} = \varepsilon_{\mathrm{p}\varphi\varphi} = -\frac{1}{2} \, p.
\ee

For a perfectly elastoplastic material, we can express the free energy and the dissipation potential as
\bea
\label{eq101a}
\Psi(u,p) = \frac{K}{2} \left( u^\prime+\frac{2}{r}u\right)^2 + \frac{2}{3}\, \mu \left( u^\prime-\frac{1}{r}u-\frac{3}{2}p \right)^2 , \quad
\Delta(\dot p) = \tau_y \left\| \dot p \right\|,
\eea
where $K$ denotes the bulk modulus.

In general, we would like to find $u(r,t)$ and $p(r,t)$ for given $u_\mathrm{in}(t)=u(r_\mathrm{in},t)$ and $u_\mathrm{out}(t)=u(r_\mathrm{out},t)$. This is in principle an infinite-dimensional problem, since the internal and external variables $p(r,t)$ and $u(r,t)$, respectively, have to be known for all values of $r$. Following a similar procedure to the $1^{st}$ example, and in order to devise a macroscopic model of the system under consideration, let us subdivide $\Omega$ into $N_\mathrm{sd}$  sub-domains 
\be
\label{eq102}
\Omega_i = \{\bfm{r} \in \mathbb{R}^3 \,|\, r_{i-1} \leq |\bfm{r}| \leq r_{i}\},
\ee
with
\be
\label{eq103}
r_\mathrm{in}=r_0 < r_1 < \ldots < r_{N_\mathrm{sd}-1} < r_{N_\mathrm{sd}}=r_\mathrm{out}.
\ee
Let us introduce local averages for the spherical body as 
\be
\label{eq104}
\langle f \rangle_i =  \frac{1}{|\Omega_i|} \int_{\Omega_i} f \dd\bfm{r} = \frac{3}{(r_i^3-r_{i-1}^3)} \, \int_{r_{i-1}}^{r_i} f r^2 \dd r.
\ee
We are interested only in the mean values of $p$ in each sub-domain. So, we introduce
\be
\label{eq105}
p_i =  \langle p \rangle_i,
\ee
which are the values that constitute our essential internal parameters,
\be
\label{eq106} 
\bfm{z}_\mathrm{ess} = \{p_i\}. 
\ee
The essential external parameters will be the boundary displacements of the spherical body under consideration
\be
\label{eq107}
\bfm{x}_\mathrm{ess} = \{u_\mathrm{in},u_\mathrm{out}\}.
\ee
There is a fundamental difference between the present problem and the torsion bar presented in Sec.\ \ref{cyl}. In the torsion example, the displacement field is completely determined by twist, i.e.\ by the essential external parameter.  In our sphere example, however, the displacement field  now depends on $p(r,t)$ and the boundary conditions. Or, to put it in the language of Sec.\ \ref{sec:vf}, for the torsion bar, the projection operator $\mathrm{P}$ defined in Eq.\ \eqref{eq5} is trivial, because there is only one external variable ($\theta^\prime$), by default being essential. Now $X=\left\lbrace u(r,t) ~\left|~ u_\mathrm{in}(t)=u(r_\mathrm{in},t), u_\mathrm{out}(t)=u(r_\mathrm{out},t) \right. \right\rbrace$, an infinite-dimensional space, and $\mathrm{P}$ has to be calculated. 

In order to find the effective energy $\Psi_\mathrm{macro}(u_\mathrm{in},u_\mathrm{out},p_1,\ldots,p_{N_\mathrm{sd}} )$, we will follow a two step strategy. Let us first consider individual subdomains and calculate intermediate energies
 \be
\label{eq109}
\Psi^i_\mathrm{int}(u_{i-1},u_i,p_i) 
 =
\inf{u,p}{\int_{r_{i-1}}^{r_i} 4\pi \, \Psi(u,p) \, r^2 \dd r}{u(r_{i-1})=u_{i-1},\,u(r_i)=u_i,\,\langle p \rangle_i=p_i}.
\ee
Then the macroscopic free energy can be calculated as 
\be
\label{eq110}
\Psi_\mathrm{macro}(u_\mathrm{in},u_\mathrm{out},p_1,\ldots,p_{N_\mathrm{sd}} ) 
 =
\inf{u_i}{\sum_{j=1}^{N_\mathrm{sd}} \Psi^j_\mathrm{int}(u_{j-1},u_j,p_j)}{u_0=u_\mathrm{in},u_{N_\mathrm{sd}}=u_\mathrm{out}}.
\ee
Similarly as in Sec.\ \ref{cyl}, we introduce Lagrange-multipliers $\lambda_i$ in order to take into account the constrains in Eq.\ \eqref{eq105}. Then variation with respect to $u$ in \eqref{eq109} gives
\be
\label{eq111}
- K \left( r^2 \left( u^\prime+\frac{2}{r}u\right) \right)^\prime + K \left( 2 r \left( u^\prime+\frac{2}{r}u\right) \right)
- \frac{4}{3}\, \mu \left( r^2 \left( u^\prime-\frac{1}{r}u-\frac{3}{2}p \right) \right)^\prime 
- \frac{4}{3}\, \mu \left( r \left( u^\prime-\frac{1}{r}u-\frac{3}{2}p \right) \right) = 0,
\ee
while variation with respect to $p$ yields
\be
\label{eq112}
- 8 \pi \mu \left( u^\prime-\frac{1}{r}u-\frac{3}{2}p \right) = \frac{\lambda_i}{m^{(1)}_i}.
\ee
Equations \eqref{eq111} and \eqref{eq112} have the solutions
\be
\label{eq113}
u = c_1 r + c_2 \frac{1}{r^2} - \frac{1}{6 K}\frac{\lambda_i}{m^{(1)}_i} \, r\ln{r},
\ee
\be
\label{eq114}
p =  - 2 c_1 \frac{1}{r^3} - \frac{1}{\pi} \, \left( \frac{1}{9 K} + \frac{1}{12 \mu} \right) \frac{\lambda_i}{m^{(1)}_i},
\ee
where $c_1, c_2$ are integration constants. 

The parameters $\lambda_i$, $c_1, c_2$ can be determined by evaluating the constraints in Eq.\ \eqref{eq105}, $u(r_{i-1})=u_{i-1}$, and $u(r_i)=u_i$. Substitiution of Eqs.\ \eqref{eq113} and \eqref{eq114} into \eqref{eq109} then gives an explicit expression for $\Psi^i_\mathrm{int}(u_{i-1},u_i,p_i)$. Finally, the determination of $\Psi_\mathrm{macro}(u_\mathrm{in},u_\mathrm{out},p_1,\ldots,p_{N_\mathrm{sd}} )$ involves solving a linear system of equations in $u_1, \ldots, u_{N_\mathrm{sd}}$.  All of these steps are algebraically straightforward, but involved. So we employed an automated computer algebra system in order to carry them out. 
At the end, however, an explicit expression for $\Psi_\mathrm{macro}$ is obtained which is a quadratic polynomial in the variables $u_\mathrm{in},u_\mathrm{out},p_1,\ldots,p_{N_\mathrm{sd}}$. Further minimization with respect to $u_\mathrm{out}$ gives the effective free energy $\Psi_\mathrm{macro}(u_\mathrm{in},p_1,\ldots,p_{N_\mathrm{sd}} )$ corresponding to a free outer boundary (and similarly for $u_\mathrm{in}$).

For completeness, explicit expressions for the macroscopic energy are given below where we use the same elastic properties as in the torsion problem, $r_\mathrm{in}=5 \,\mathrm{mm}$ and $r_\mathrm{out}=10\, \mathrm{mm}$.  Two cases are shown, $N_\mathrm{sd}=2$ sub-domains and  $N_\mathrm{sd}=5$ sub-domains:
\be
\label{eq110a}
\Psi_\mathrm{macro}(u_\mathrm{in},p_1,p_2) 
= \sum_{|\alpha|=2} c_\alpha u_\mathrm{in}^{\alpha_0} p_1^{\alpha_1} p_2^{\alpha_2}\,,
\qquad
\Psi_\mathrm{macro}(u_\mathrm{in},p_1,\ldots,p_5) 
= \sum_{|\alpha|=2} c_\alpha u_\mathrm{in}^{\alpha_0} p_1^{\alpha_1} p_2^{\alpha_2}
p_3^{\alpha_3}p_4^{\alpha_4}p_5^{\alpha_5}
\,,
\ee
where the corresponding multi-indices are given in Tables \ref{t:multii2} and \ref{t:multii5}.

\begin{table}\centering
\caption{Multi-indices for 2 sub-domain case}\label{t:multii2}
\begin{tabular}{l rrrrrr}
\hline
$\alpha$& (2,0,0) & (0,2,0) & (0,0,2) & (1,1,0) & (1,0,1) & (0,1,1) \\
$c_\alpha$&2674.18 & 47004.4 & 71293 & 20744.1 & 14718.2 & 19892.3\\
\hline
\end{tabular}
\end{table}
\begin{table}\centering
\caption{Multi-indices for 5 sub-domain case}\label{t:multii5}
\begin{tabular}{l rrrrrr}
\hline
$\alpha$& (2,0,0,0,0,0) & (0,2,0,0,0,0) & (0,0,2,0,0,0) & (0,0,0,2,0,0) & (0,0,0,0,2,0) & (0,0,0,0,0,2) \\
$c_\alpha$&3082.72 & 13096.4 & 16245.2 & 20379.1& 25357 & 31108.1\\
$\alpha$& (1,1,0,0,0,0) &(1,0,1,0,0,0) &(1,0,0,1,0,0) &(1,0,0,0,1,0) &(1,0,0,0,0,1)& \\
$c_\alpha$&9824.36  & 8306.38 & 7195.31 & 6446.71 & 5677.33& \\
$\alpha$& (0,1,1,0,0,0) &(0,1,0,1,0,0) &(0,1,0,0,1,0) &(0,1,0,0,0,1) && \\
$c_\alpha$&5048.11 & 4372.87& 3857.14 & 3450.33 & &  \\
$\alpha$& (0,0,1,1,0,0)& (0,0,1,0,1,0)&(0,0,1,0,0,1) &(0,0,0,1,1,0) &(0,0,0,1,0,1) &(0,0,0,0,1,1)  \\
$c_\alpha$&3697.21& 3261.17 & 2917.21& 2824.95 & 2527.01 & 2228.98 \\
\hline
\end{tabular}
\end{table}
Employing the same argument as in Sec.\ \ref{cyl}, we obtain the effective dissipation potential
\be
\label{eq115}
\Delta_\mathrm{macro}(\dot p_1,\ldots,\dot p_{N_\mathrm{sd}} ) 
= \sum_{j=1}^{N_\mathrm{sd}}\,2\pi \, m^{(2)}_j \tau_y \, \vert \dot p_j \vert.
\ee
Introducing thermodynamically conjugate driving forces
\be
\label{eq116}
f_i = - \frac{\partial \Psi_\mathrm{macro}}{\partial p_i},
\ee
The effective Biot-equations are given as
\be
\label{eq117}
f_i \in 2\pi \, m^{(2)}_i \tau_y \, \mathrm{sign} \, \dot p_i, \quad \text{ for } i = 1,\ldots,N_\mathrm{sd}.
\ee
The corresponding yield condition and flow rule are established as
\be
\label{eq118}
\left| f_i \right| \le 2\pi \, m^{(2)}_i \tau_y, \quad \dot p_i = \lambda_i \, f_i, \quad \lambda_i \ge 0, \quad \text{ for } i = 1,\ldots,N_\mathrm{sd}.
\ee
Note that in contrast to the torsion problem in Sec.\ \ref{cyl}, the parameters $f_i$ depend on all external and internal essential variables. This means, that the yield conditions in Eq.\ \eqref{eq118} are coupled resulting in a problem of \textit{multi-surface plasticity}. For the examples below, we will solve this in a staggered scheme, by evaluating the yield condition involving $f_i$ for the individual $p_i$ keeping all the other variables $p_j$ fixed. This approach is convenient for demonstration purposes and accurate for small time-steps.

As noted, for our computations we use the same material parameters as in the torsion problem. For the sphere geometry, the internal and external radii are $r_\mathrm{in}=5 \,\mathrm{mm}$ and $r_\mathrm{out}=10 \,\mathrm{mm}$.
We vary the outer displacement as $u_\mathrm{out} = A_2 g(t)$, see Fig.~\ref{fig0b}, with $A_2= 3\, \mathrm{mm}$ and $t_0 = \frac{0.5\times 5}{600}\, \mathrm{sec}$; further we fix the inner boundary,  $u_\mathrm{in}=0$. 
Beside the effective model computations, a converged finite element solution is determined, so that a comparison between the results can be investigated.

\begin{figure}
\centering
  \includegraphics[width=\textwidth]{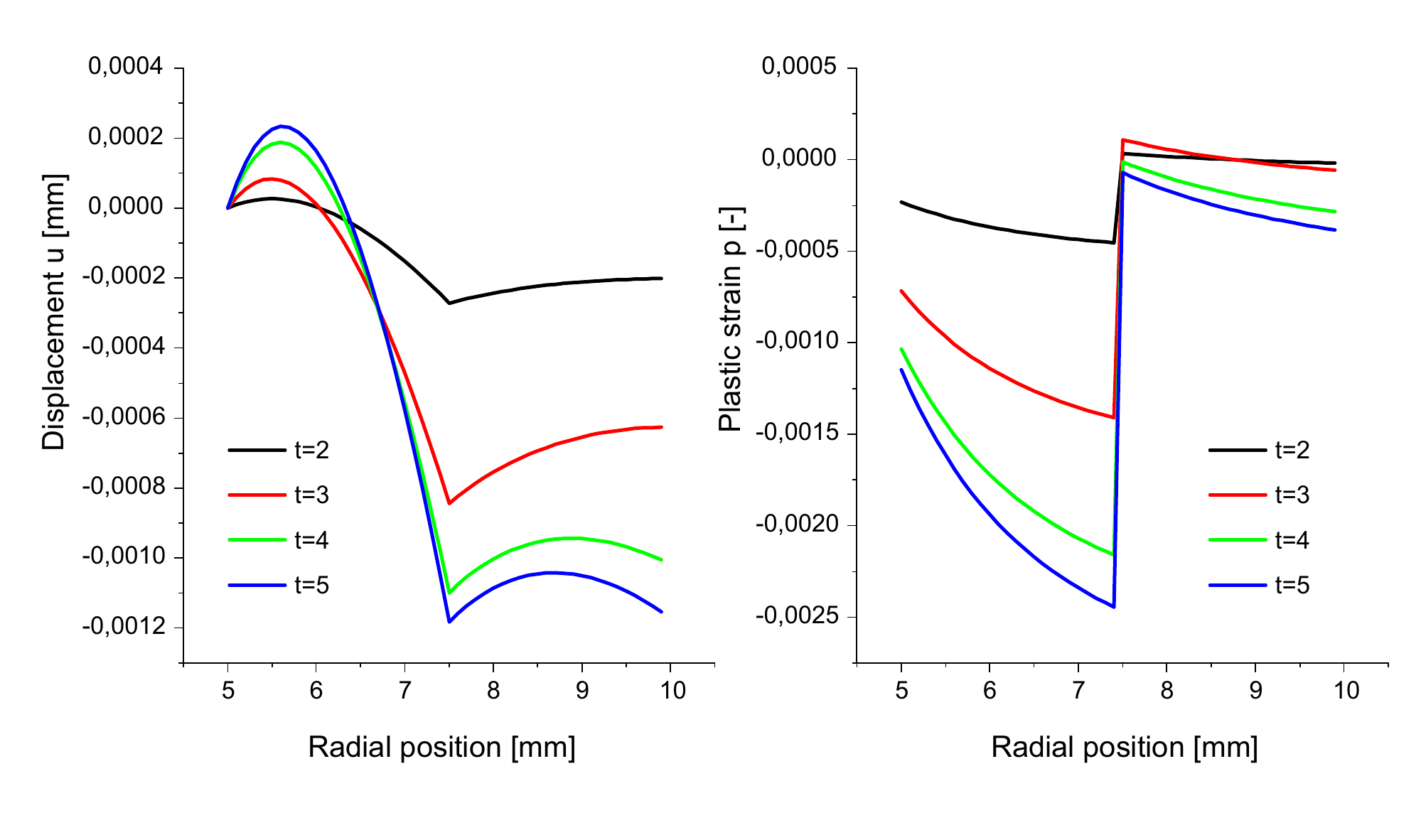}  
    \caption{Illustration of the displacement field (left) and plastic strains (right) versus radial position between the inner and outer radius for times $t=2, 3, 4, 5~\mathrm{sec}$ and 2 sub-domains.}
    \label{fig0f}
 \end{figure}

We examine the displacement and plastic strain behavior for 2 sub-domains at specific times $t=2, 3, 4, 5~\mathrm{sec}$ as shown in  Fig.~\ref{fig0f}.
The plastic strains show a jump from negative to positive values at the sub-domain interface. As in the torsion model, the recovered plastic strains are non-constant within the sub-domains, as can be deduced from Eq.\ \eqref{eq114}.

\begin{figure}
\centering
  \includegraphics[width=\textwidth]{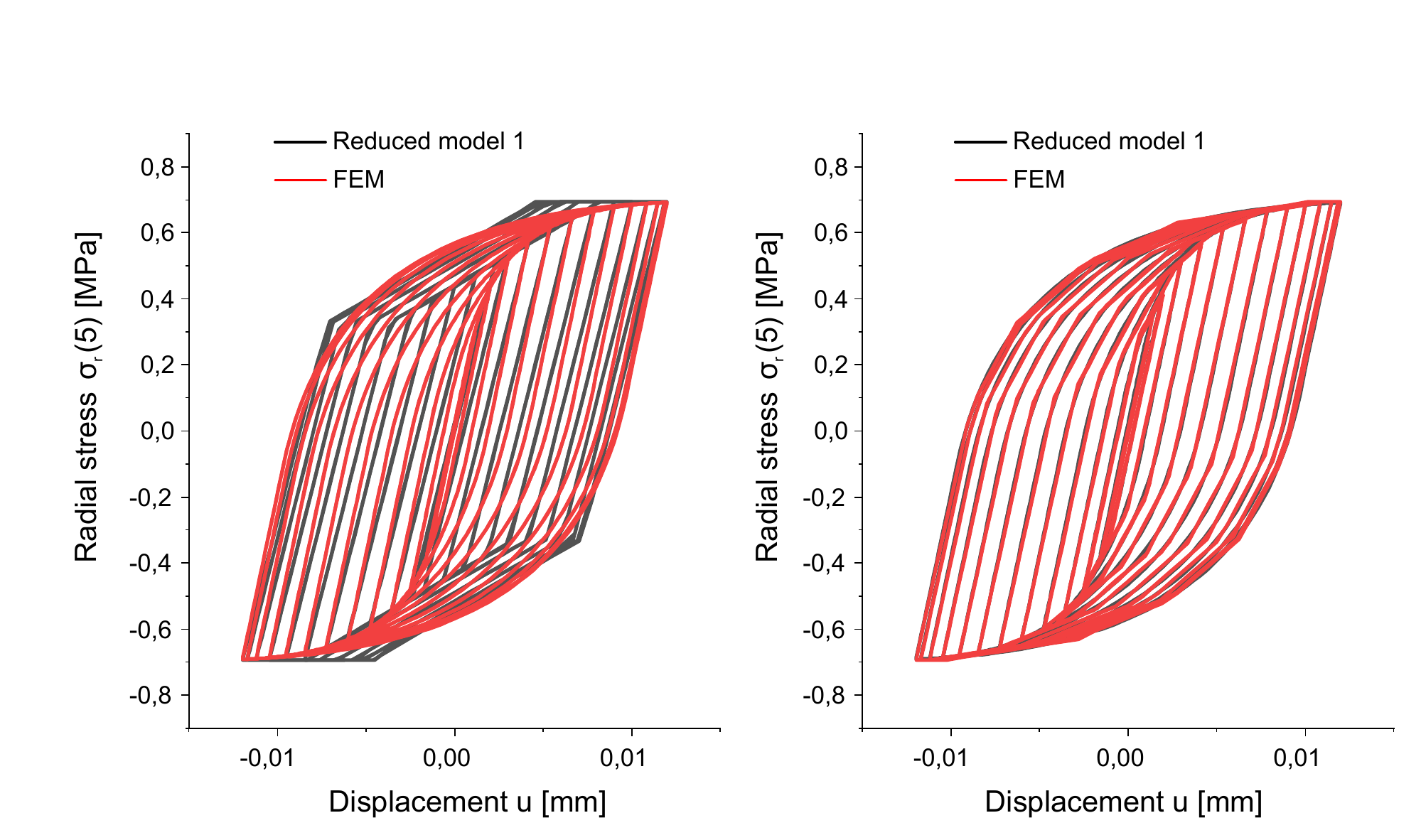}  
    \caption{Comparison between the radial stress on the inner boundary versus the radial displacement 
    on the outer boundary from the reduced model and the finite element model, once for 2 sub-domains (left) and once for 5 sub-domains(right).}
    \label{fig0g}
 \end{figure}

In Fig.~\ref{fig0g}, we see a comparison of between the effective model and a converged finite element computation.  Shown is the response of the radial stress on the inner boundary versus the imposed external displacement, once for 2 sub-domains and once for 5 sub-domains. We observe the good agreement between the effective model and the converged numerical solution, which is also getting quantitatively and qualitatively better as the number of sub-domains increases.

\section{Homogenization of elastoplastic materials with periodic microstructure}

\subsection{The homogenization problem}

Let us consider now a material having a microstructure defined by periodically varying $\mathbb{C}(\bfm{r})$ and $\sigy(\bfr)$ over a representative volume element $\Omega$, which we take to be rectangular. The average of a quantity $f$ over $\Omega$ is given by
\be
\label{eq14a}
\langle f \rangle =  \frac{1}{|\Omega|} \int_{\Omega} f \dd\bfm{r}.
\ee
The macroscopic material behavior may be determined by subjecting the representative volume element to a macroscopic strain $\bfm{e}_\mathrm{M}=\bfm{e}_\mathrm{M}(t)=\langle \beps \rangle$ and calculating the macroscopic stress $\bfsigma_\mathrm{M}=\bfsigma_\mathrm{M}(t)=\langle \bfsigma \rangle$. The imposed macroscopic strain may be realized by decomposing the displacement field as
\be
\label{eq14b}
\bfm{u} = \bfm{e}_\mathrm{M}\cdot\bfm{r} + \bfm{u}_\mathrm{per},
\ee
where $\bfm{u}_\mathrm{per}$ satisfies so-called periodic boundary conditions, i.e.\ $\bfm{u}_\mathrm{per}(\bfm{r}_-) = \bfm{u}_\mathrm{per}(\bfm{r}_+)$ on corresponding boundary points $\bfm{r}_-$, $\bfm{r}_+$ on opposite sides of $\Omega$.

\subsection{The macroscopic model and essential variables}

The problem of determination of the macroscopic model for such as system as outlined in Sec.\ \ref{sec:3} is in principle infinite-dimensional because the plastic strains $\bepsp(\bfm{r},t)$ have to be known at all points of $\Omega$ for all times. In order to devise a macroscopic model of the system under consideration let us subdivide our representative volume element into $N_\mathrm{sd}$ distinct sub-domains $\Omega_i$ such that
\be
\label{eq15}
\Omega = \bigcup_{i=1}^{N_\mathrm{sd}} \Omega_i,
\ee
and introduce local averages as 
\be
\label{eq15a}
\langle f \rangle_i =  \frac{1}{|\Omega_i|} \int_{\Omega_i} f \dd\bfm{r}.
\ee
In every sub-domain we are only interested in mean values of the plastic strains.
Hence, we introduce
\be
\label{eq16}
\bfm{e}_{\mathrm{p}i} =  \langle \bepsp \rangle_i,
\ee
and choose those values to constitute our essential internal parameters,
\be
\label{eq16a} 
\bfm{z}_\mathrm{ess} = \bfm{e}_{\mathrm{p}} = \left( \bfm{e}_{\mathrm{p}i}  \right). 
\ee
As essential external parameters we will choose the macroscopic strain,
\be
\label{eq17}
\bfm{x}_\mathrm{ess} = \bfm{e}_\mathrm{M}.
\ee

\subsection{Macroscopic free energy}

With the definitions above the macroscopic free energy is given as
\be
\label{eq18}
\Psi_\mathrm{macro}(\bfm{e}_\mathrm{M},\bfm{e}_{\mathrm{p}} ) =
\inf{\bfm{u}_\mathrm{per},\, \bepsp}{\left\langle \frac{1}{2} \left( \beps - \bepsp \right) : \mathbb{C} : \left( \beps - \bepsp \right) \right\rangle}
{\bfm{u} = \bfm{e}_\mathrm{M}\cdot \bfm{r} + \bfm{u}_\mathrm{per}, \quad \bfm{e}_{\mathrm{p}i} =  \langle \bepsp \rangle_i }.
\ee
Introducing Lagrange-parameters $\bfmu_{i}$ for the second constraint we obtain the Lagrangian for the minimization problem above as
\be
\label{eq20}
L =
\left\langle \frac{1}{2} \left( \beps - \bepsp \right) : \mathbb{C} : \left( \beps - \bepsp \right) \right\rangle + \sum_{i=1}^{N_\mathrm{sd}} \bfmu_{i} : \left( \bfm{e}_{\mathrm{p}i} - \langle \bepsp \rangle_i \right).
\ee
As stationarity conditions with respect to $\bfm{u}_\mathrm{per}$ and $\bfm{e}_{\mathrm{p}i}$, respectively, we obtain
\be
\label{eq21}
\nabla\cdot \left( \mathbb{C} : \left( \beps - \bepsp \right) \right) =\mathbf{0} \quad \mbox{in} \; \Omega,
\ee
\be
\label{eq22}
\mathbb{C} : \left( \beps - \bepsp \right) = \bfmu_{i} \quad \mbox{in} \; \Omega_i.
\ee
From Eq.\  \eqref{eq22} we see that the stress $\bfsigma=\bfmu_i$ is constant in every sub-domain $\Omega_i$ and that condition \eqref{eq21} is trivially satisfied. 

Averaging in Eq.\  \eqref{eq22} gives
\be
\label{eq23}
\bfm{e}_i - \bfm{e}_{\mathrm{p}i} = \mathbb{C}_{\mathrm{eff}i}^{-1} : \bfmu_{i},
\ee
and thus
\be
\label{eq24}
\beps - \bepsp = \mathbb{C}^{-1} : \mathbb{C}_{\mathrm{eff}i} : ( \bfm{e}_i - \bfm{e}_{\mathrm{p}i} ),
\ee
where we introduced the abbreviations
\be
\label{eq25}
\bfm{e}_i = \langle \beps \rangle_i, \quad \mathbb{C}_{\mathrm{eff}i} = \left( \langle 
\mathbb{C}^{-1} \rangle_i \right)^{-1}.
\ee
Substitution of Eq.\  \eqref{eq24} into \eqref{eq18} gives
\be
\label{eq26}
\Psi_\mathrm{macro}(\bfm{e}_\mathrm{M},\bfm{e}_{\mathrm{p}} ) =
\inf{\bfm{u}_\mathrm{per}}{\sum_{i=1}^{N_\mathrm{sd}} \frac{|\Omega_i|}{|\Omega|} \, \frac{1}{2} \left( \bfm{e}_i - \bfm{e}_{\mathrm{p}i} \right) : \mathbb{C}_{\mathrm{eff}i} : \left( \bfm{e}_i - \bfm{e}_{\mathrm{p}i} \right)}
{\bfm{u} = \bfm{e}_\mathrm{M}\cdot \bfm{r} + \bfm{u}_\mathrm{per}}.
\ee
It can be seen that $\Psi_\mathrm{macro}(\bfm{e}_\mathrm{M},\bfm{e}_{\mathrm{p}} )$ is a quadratic expression in $\bfm{e}_\mathrm{M}$ and $\bfm{e}_{\mathrm{p}}$.

From Eq.\  \eqref{eq26} we obtain immediately the conjugate driving forces to the essential internal parameters as
\be
\label{eq26a}
\bfm{q}_i = -\frac{\partial \Psi_\mathrm{macro}}{\partial \bfm{e}_{\mathrm{p}i}} = \frac{|\Omega_i|}{|\Omega|} \, \bfsigma_i,
\ee
where
\be
\label{eq26b}
\bfsigma_i = \mathbb{C}_{\mathrm{eff}i} : \left( \bfm{e}_i - \bfm{e}_{\mathrm{p}i} \right)
\ee
denotes the constant stress in every sub-domain.

\subsection{Macroscopic dissipation potential}

The macroscopic dissipation potential is defined as
\be
\label{eq42}
\Delta_\mathrm{macro}(\dot{\bfm{e}}_{\mathrm{p}}) = 
\inf{\dbepsp}{\left\langle\sqrt{2/3} \, \sigy  \left\| \dbepsp \right\| \right\rangle}{ \dot{\bfm{e}}_{\mathrm{p}i} =  \langle \dbepsp \rangle_i }.
\ee
Because $\dbepsp$ occurs in Eq.\  \eqref{eq42} in an algebraic way only, the dissipation potential can immediately be split into contributions defined on the sub-domains:
\be
\label{eq43}
\Delta_\mathrm{macro}(\dot{\bfm{e}}_{\mathrm{p}}) = \frac{1}{|\Omega|} \sum_{i=1}^{N_\mathrm{sd}} |\Omega_i| \, \Delta_\mathrm{macro}^i(\dot{\bfm{e}}_{\mathrm{p}i}),
\ee
with
\be
\label{eq44}
\Delta_\mathrm{macro}^i(\dot{\bfm{e}}_{\mathrm{p}i}) = 
\inf{\dbepsp}{\left\langle\sqrt{2/3} \, \sigy  \left\| \dbepsp \right\| \right\rangle_i}{ \dot{\bfm{e}}_{\mathrm{p}i} =  \langle \dbepsp \rangle_i }.
\ee
In general, $\Delta_\mathrm{macro}^i(\dot{\bfm{e}}_{\mathrm{p}i})$ can only be calculated numerically. For this reason, let us assume that $\sigy = \sigyi$ is constant in every sub-domain. Then it immediately follows that
\be
\label{eq45}
\Delta_\mathrm{macro}^i(\dot{\bfm{e}}_{\mathrm{p}i}) = \sqrt{2/3} \,\sigyi \, \|\dot{\bfm{e}}_{\mathrm{p}i}\|
\ee
and we obtain
\be
\label{eq46}
\Delta_\mathrm{macro}(\dot{\bfm{e}}_{\mathrm{p}}) = \sqrt{2/3} \frac{1}{|\Omega|} \sum_{i=1}^{N_\mathrm{sd}} |\Omega_i| \, \sigyi \, \|\dot{\bfm{e}}_{\mathrm{p}i}\|.
\ee
In the light of Eqs.\  \eqref{eq26a} and \eqref{eq46} we see that we have formally decoupled yield conditions of the form
\be
\label{eq47}
\| \dev \,\bfsigma_i \| \leq \sqrt{2/3} \, \sigyi
\ee
in every sub-domain. However, one should keep in mind that $\bfsigma_i$ depends on all essential internal variables $\bfm{e}_{\mathrm{p}} = \left( \bfm{e}_{\mathrm{p}i}  \right)$.

\subsection{Polyhedral sub-domains}

Let us consider now microstructures consisting of polyhedral sub-domains bounded by facets $F_i$. Of course, any microstructure may be approximated in this way with arbitrary accuracy by choosing the facets small enough. Let the facets have outward normal vectors $\bfm{n}_i$. Then the average strains in every sub-domain can be calculated according to
\be
\label{eq48}
\bfm{e}_i = \bfm{e}_\mathrm{M} + \frac{1}{|\Omega_i|} \, \sum_{F_j \subset \partial\Omega_i} \sym  (\bfm{a}_j \otimes \bfm{n}_j),
\ee
where $\sym(\bfm{T})=\frac{1}{2}(\bfm{T}+\bfm{T}^\mathrm{T})$, and the amplitude vectors $\bfm{a}_i$ are given by the surface integrals
\be
\label{eq49}
\bfm{a}_i = \int_{F_i} \bfm{u}_\mathrm{per} \dd S.
\ee
Note that because of the periodic boundary conditions some of the $\bfm{a}_j$ are dependent if $F_j \subset \partial\Omega$. Let the number of independent amplitude vectors be $N_\mathrm{a}$. 

The macroscopic free energy can now be written as
\be
\label{eq49a}
\Psi_\mathrm{macro}(\bfm{e}_\mathrm{M},\bfm{e}_{\mathrm{p}} ) =
\infr{\bfm{a}_i,\ldots,\bfm{a}_{N_\mathrm{a}}}
{\Psi_\mathrm{rve}(\bfm{e}_\mathrm{M},\bfm{e}_{\mathrm{p}},\bfm{a}_1,\ldots,\bfm{a}_{N_\mathrm{a}} )}{}.
\ee
where
\be
\label{eq50}
\Psi_\mathrm{rve}(\bfm{e}_\mathrm{M},\bfm{e}_{\mathrm{p}},\bfm{a}_1,\ldots,\bfm{a}_{N_\mathrm{a}} ) =
\sum_{i=1}^{N_\mathrm{sd}} \frac{|\Omega_i|}{|\Omega|} \, \frac{1}{2} \left( \bfm{e}_i - \bfm{e}_{\mathrm{p}i} \right) : \mathbb{C}_{\mathrm{eff}i} : \left( \bfm{e}_i - \bfm{e}_{\mathrm{p}i} \right),
\ee
where $\bfm{e}_i$ is given by Eq.~\eqref{eq48} and it is recalled that $\bfm{e}_\mathrm{p} = (\bfm{e}_{\mathrm{p}i})$.
Interestingly, $\Psi_\mathrm{macro}$ is only dependent on the averages of the displacement field $\bfm{u}_\mathrm{per}$ over the facets. Thus we have generated a theory involving only averages over the sub-domains of the various quantities. However, neither $\beps$ nor $\bepsp$ are required to be constant on the sub-domains.

The stationarity conditions corresponding to minimization are
\be
\label{eq50a}
\frac{\partial \Psi_\mathrm{rve}}{\partial \bfm{a}_i} =0, \qquad i=1,\ldots,N_\mathrm{a}.
\ee
The system of equations \eqref{eq50a} is underdetermined, i.e.\ has multiple solutions. However, all solutions give the same minimum energy $\Psi_\mathrm{macro}$. The macroscopic energy will be quadratic of the form
\be
\label{eq50b}
\Psi_\mathrm{macro} = \frac{1}{2} \, \bfm{e}_\mathrm{M}:\mathbb{C}_\mathrm{eff}:\bfm{e}_\mathrm{M} - \sum_{i=1}^{N_\mathrm{sd}} \bfm{e}_{\mathrm{p}i}:\mathbb{F}_{\mathrm{eff}i}:\bfm{e}_\mathrm{M} + \sum_{i,j=1}^{N_\mathrm{sd}} \frac{1}{2} \, \bfm{e}_{\mathrm{p}i}:\mathbb{G}_{\mathrm{eff}ij}:\bfm{e}_{\mathrm{p}j}.
\ee
The conjugate driving forces follow as
\be
\label{eq50c}
\bfm{q}_i = -\frac{\partial \Psi_\mathrm{macro}}{\partial \bfm{e}_{\mathrm{p}i}} = \frac{|\Omega_i|}{|\Omega|} \, \bfsigma_i = \mathbb{F}_{\mathrm{eff}i}:\bfm{e}_\mathrm{M} - \sum_{j=1}^{N_\mathrm{sd}} \mathbb{G}_{\mathrm{eff}ij}:\bfm{e}_{\mathrm{p}j}.
\ee
The dissipation potential and yield conditions are not affected and still given by Eqs.\  \eqref{eq46} and \eqref{eq47}.

\subsection{A basic example}

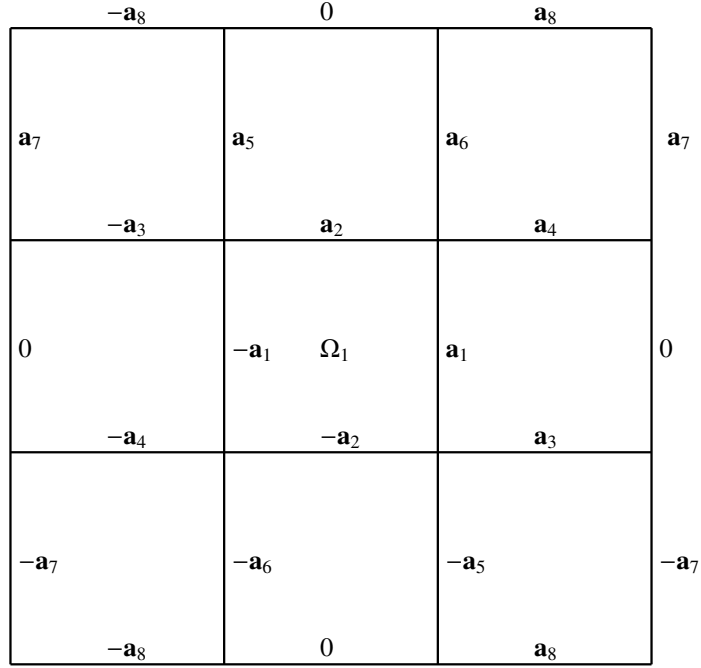
\begin{figure}[h]
    \centering
    \begin{picture}(240,260)
    \linethickness{0.3mm}
    \put(0,0){\line(1,0){240}}
    \put(0,80){\line(1,0){240}}
    \put(0,160){\line(1,0){240}}
    \put(0,240){\line(1,0){240}}
    \put(0,0){\line(0,1){240}}
    \put(80,0){\line(0,1){240}}
    \put(160,0){\line(0,1){240}}
    \put(240,0){\line(0,1){240}}
    \put(116,116){$\Omega_1$}
    \put(163,116){$\bfm{a}_1$}
    \put(83,116){$-\bfm{a}_1$}
    \put(116,163){$\bfm{a}_2$}
    \put(116,83){$-\bfm{a}_2$}
    \put(243,116){$0$}
    \put(3,116){$0$}
    \put(116,243){$0$}
    \put(116,3){$0$}
    \put(196,163){$\bfm{a}_4$}
    \put(196,83){$\bfm{a}_3$}
    \put(36,163){$-\bfm{a}_3$}
    \put(36,83){$-\bfm{a}_4$}
    \put(163,196){$\bfm{a}_6$}
    \put(83,196){$\bfm{a}_5$}
    \put(163,36){$-\bfm{a}_5$}
    \put(83,36){$-\bfm{a}_6$}
    \put(246,196){$\bfm{a}_7$}
    \put(3,196){$\bfm{a}_7$}
    \put(243,36){$-\bfm{a}_7$}
    \put(3,36){$-\bfm{a}_7$}
    \put(196,243){$\bfm{a}_8$}
    \put(196,3){$\bfm{a}_8$}
    \put(36,243){$-\bfm{a}_8$}
    \put(36,3){$-\bfm{a}_8$}
    \end{picture}
    \caption{Representative volume element with sub-domains and amplitude vectors for symmetric case, with $\mathrm{ Mat_1}$ assigned to sub-domain $\Omega_1$ and $\mathrm{ Mat_2}$ elsewhere.}
    \label{fig1}
\end{figure}

As a concrete example, consider  a domain, a representative volume element, shown in Fig.~\ref{fig1}, consisting of nine sub-domains. Within the central sub-domain $\Omega_1$ we have constant elasticity tensor $\mathbb{C}_{1}$ and yield stress $\sigma_\mathrm{y1}$, elsewhere, in the surrounding material, we have constant elasticity tensor $\mathbb{C}_{2}$ and yield stress $\sigma_\mathrm{y2}$.

Periodic boundary conditions require, that only five of the sub-domains possess different average strains and that only eight different amplitude vectors have to be considered.
Let the volumes of the sub-domains be normalized as $|\Omega_i|=1$, and let $\bfm{n}_1=(1,0)$, $\bfm{n}_2=(0,1)$ be two normal vectors.
According to Eq.\  \eqref{eq48} the average total strains in the sub-domains are given as
\be
\begin{array}{l}
\label{eq51}
\D \bfm{e}_1 = \bfm{e}_\mathrm{M} + 2 \, \sym \left(\bfm{a}_1 \otimes \bfm{n}_1 + \bfm{a}_2 \otimes \bfm{n}_2\right), \\
\D \bfm{e}_2 = \bfm{e}_\mathrm{M} + \sym \left(-\bfm{a}_1 \otimes \bfm{n}_1 + (\bfm{a}_4 - \bfm{a}_3) \otimes \bfm{n}_2\right), \\
\D \bfm{e}_3 = \bfm{e}_\mathrm{M} + \sym \left((\bfm{a}_6 - \bfm{a}_5) \otimes \bfm{n}_1 - \bfm{a}_2 \otimes \bfm{n}_2\right), \\
\D \bfm{e}_4 = \bfm{e}_\mathrm{M} + \sym \left((\bfm{a}_7 - \bfm{a}_6) \otimes \bfm{n}_1 + (\bfm{a}_8 - \bfm{a}_4) \otimes \bfm{n}_2\right), \\
\D \bfm{e}_5 = \bfm{e}_\mathrm{M} + \sym \left((\bfm{a}_5 - \bfm{a}_7) \otimes \bfm{n}_1 + (\bfm{a}_3 - \bfm{a}_8) \otimes \bfm{n}_2\right).
\end{array}
\ee
The macroscopic free energy becomes
\be
\label{eq52}
\Psi_\mathrm{macro}(\bfm{e}_\mathrm{M},\bfm{e}_{\mathrm{p}} ) =
\infr{\bfm{a}_1,\ldots,\bfm{a}_8}{\Psi_\mathrm{rve}(\bfm{e}_\mathrm{M},\bfm{e}_{\mathrm{p}},\bfm{a}_1,\ldots,\bfm{a}_8 )}.
\ee
where
\be
\label{eq53}
\Psi_\mathrm{rve}(\bfm{e}_\mathrm{M},\bfm{e}_{\mathrm{p}},\bfm{a}_1,\ldots,\bfm{a}_8 ) =
\frac{1}{18} \, \left( \bfm{e}_1 - \bfm{e}_{\mathrm{p}1} \right) : \mathbb{C}_{1} : \left( \bfm{e}_1 - \bfm{e}_{\mathrm{p}1} \right) + \sum_{i=2}^5 \frac{1}{9} \, \left( \bfm{e}_i - \bfm{e}_{\mathrm{p}i} \right) : \mathbb{C}_{2} : \left( \bfm{e}_i - \bfm{e}_{\mathrm{p}i} \right).
\ee

\subsubsection{Numerical results}

\begin{figure}
\centering
  \includegraphics[scale=0.62]{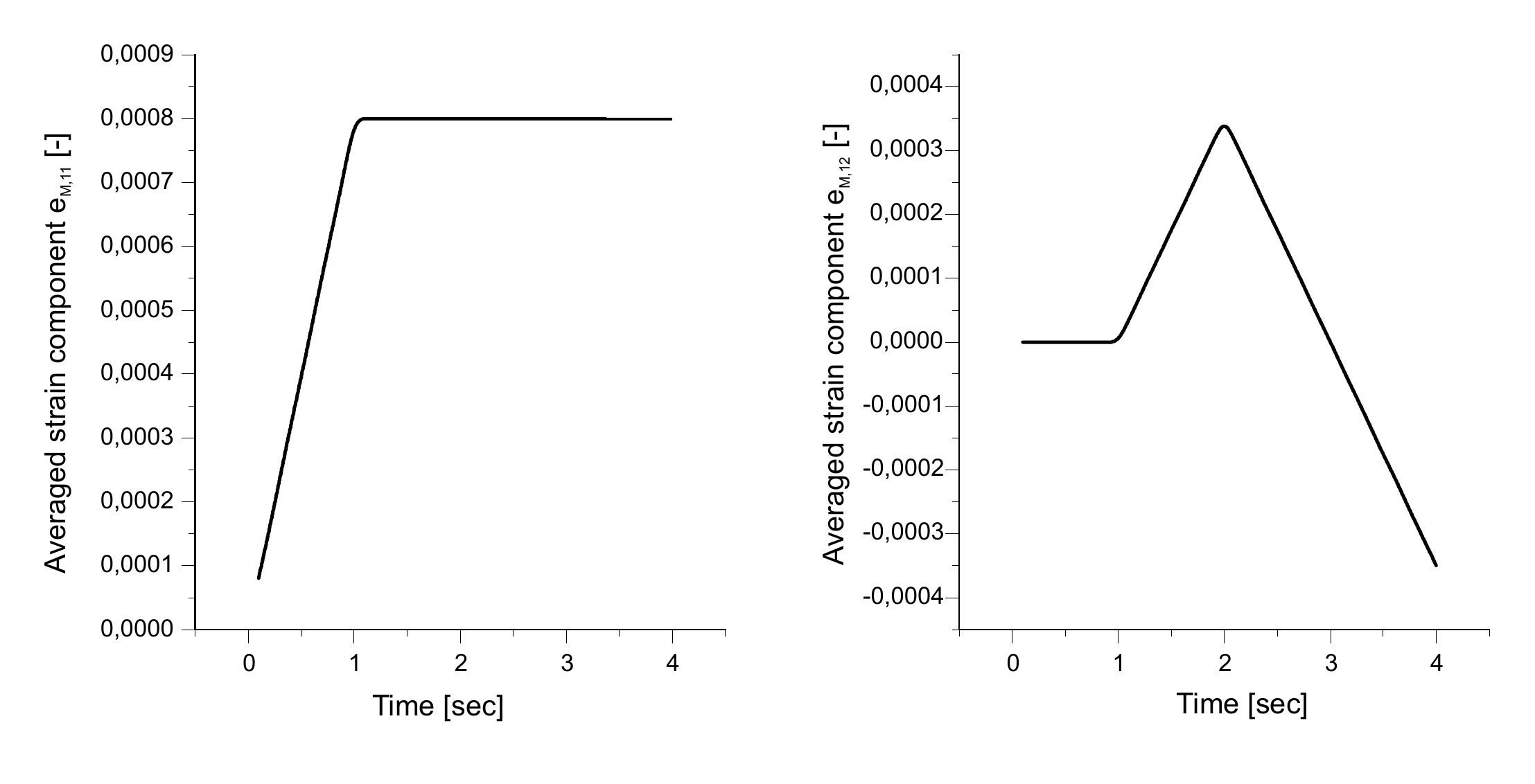}  
    \caption{Driven essential values of the non-zero components of $\bfm{e}_\mathrm{M}$, averaged strains in 1 direction (left) and the 1-2 shear (right).}
    \label{fig2}
 \end{figure}

To examine the performance of the reduced model, we
assume isotropic materials
where Mat$_\mathrm{1}$ has properties $E_1 = 1000 \,\mathrm{N/mm^{2}}$, $\nu_1=0.25$, $\sigma_{\mathrm{y}1} = 1.5 \,\mathrm{N/mm^{2}}$ in $\Omega_1$ and Mat$_\mathrm{2}$ has properties $E_2 = 5000 \,\mathrm{N/mm^{2}}$, $\nu_2=0.15$, $\sigma_{\mathrm{y}2} = 3.75 \,\mathrm{N/mm^{2}}$ elsewhere and consider a time varying loading $\bfm{e}_\mathrm{M}(t)$ consisting of a strain in the 1-direction, followed by an oscillatory shear strain in the 1-2 plane as depicted in Fig.~\ref{fig2}.

We solved the reduced model in Mathematica by minimizing the RVE free energy given in Eq.\ \eqref{eq53} with respect to the amplitude vectors, then computing the resulting macroscopic stresses by derivation of the macroscopic energy with respect to the averaged strains, and then solving the resulting equations for the plastic strains by applying the yield conditions given in Eq.\ \eqref{eq47}, which amounts to solving the material model on the micro level, locally.  To test the quality of the reduced model, its predictions are compared to the results from a much higher fidelity finite element computation, so that each sub-domain was split into 40x40 elements implying a total number of elements of 14400 to capture the micro scale. We employed quadratic quadrilateral elements (9-Nodes elements) representing the microstructure leading to 115198 degrees of freedom, and the macro behavior is obtained via the built-in Hill-mandel homogenization scheme in FEAP \citep{feapu:20} -- the FEA software we have used.  While a convergence study of the FEA computation was not explicitly made, the mesh is rather fine and represents a model with roughly three orders of magnitude, more degrees of freedom.  The computations assume plane strain conditions.

\begin{figure}
\centering
  \includegraphics[scale=0.67]{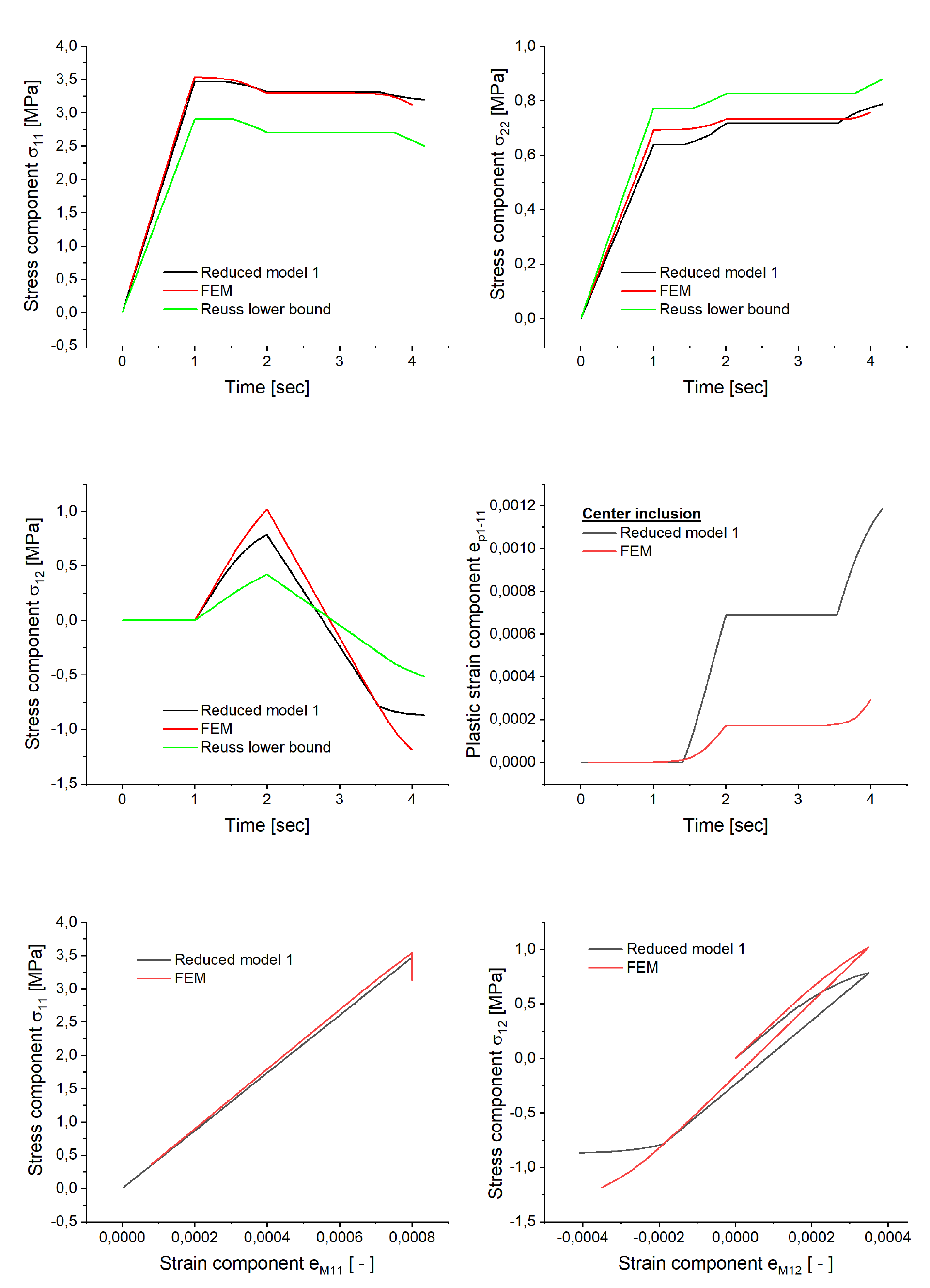}  
    \caption{Comparison of the effective model, a Reuss model, and a finite element model for a single inclusion RVE. 
    (top row and middle left pane)  macroscopic stresses versus time.  (middle right pane) 11 component of the average plastic strain, $\bfm{e}_\mathrm{p1}$, of the center inclusion versus time.  (bottom row) stress-strain response for the normal stress and strain in the 1-direction and the shear stress and strain in the 1-2 plane.}
    \label{fig3}
    \end{figure}

Figure \ref{fig3} (top row and middle left pane) shows a comparison between the macroscopic stresses versus time resulting from the effective model, a Reuss lower bound computation, and the finite element model.  Fig.~\ref{fig3} (middle right pane) shows the 11 component of the average plastic strain, $\bfm{e}_{p1}$, of the center inclusion versus time.  Lastly, Fig.~\ref{fig3} (bottom row) shows the stress-strain response for the normal stress and strain in the 1-direction and the shear stress and strain in the 1-2 plane.
The Reuss computations are performed by changing the constraint in Eq.~\eqref{eq26} to
\be
\label{eq55}
\bfm{e}_\mathrm{M}  = \bigcup_{i=1}^{N_\mathrm{sd}} \ \frac{|\Omega_i|}{|\Omega|} \, \bfm{e}_i\,,
\ee
which then ignores the symmetric gradient character of the strain field.

It can be seen that the reduced model and the FEA macroscopic stress responses versus time are in general agreement upon this path in macroscopic strain space, as is the Reuss result.  However there are, as expected, discrepancies in the stiffness of the response, especially post yield.
Nevertheless, it can be seen that the assumptions used in the reduced model were still able to capture the behavior of the plastic strains as well, but the values are higher in comparison to the finite element computations.  Strikingly, the 11 component of the mean plastic strain in the center inclusion is poorly captured by the reduced model but this appears to have only a small influence on  macroscopic stress history.

\subsection{Non-symmetric RVE example}

\begin{figure}[h]
\centering
  \includegraphics[scale=0.4]{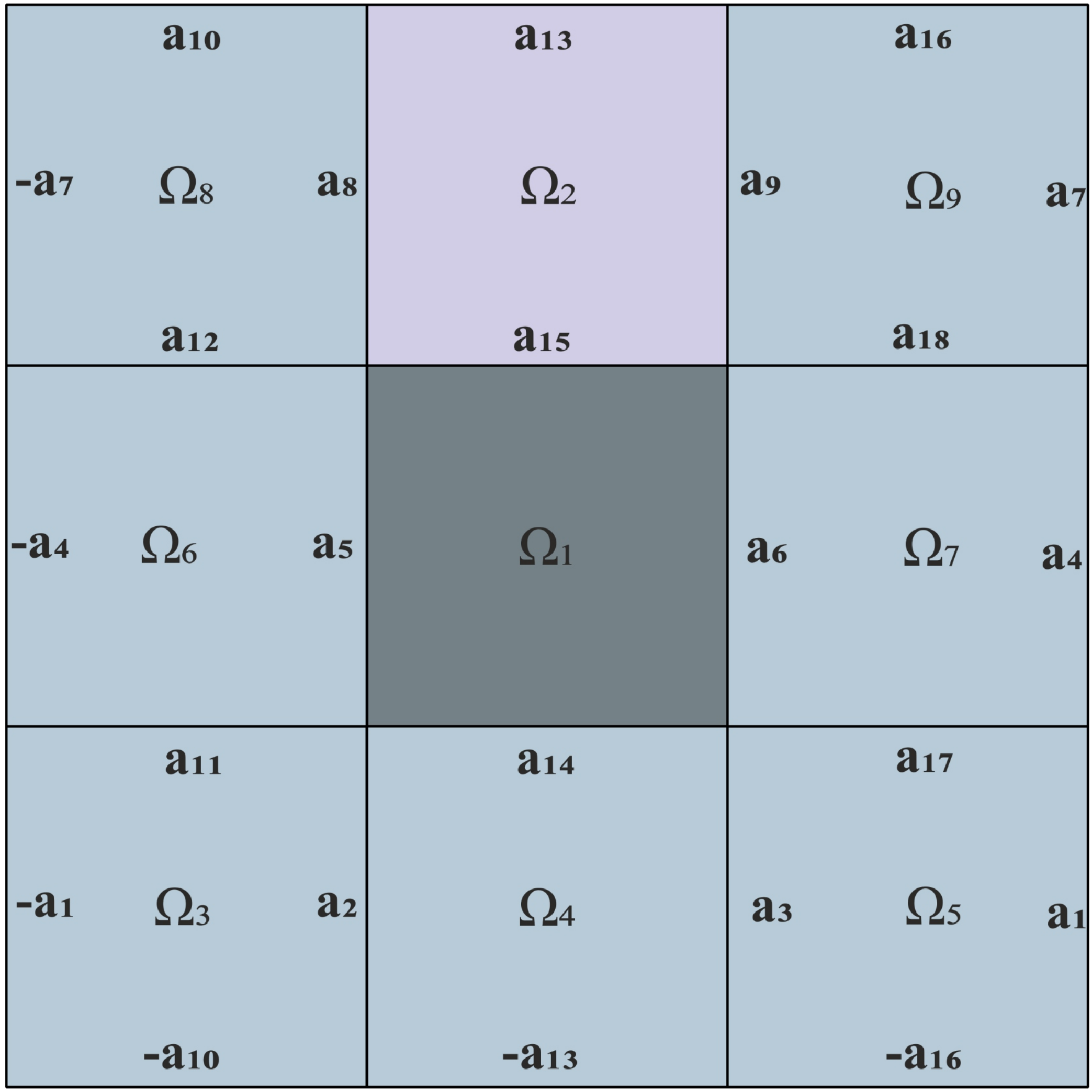}  
    \caption{representative volume element with sub-domains and amplitude vectors for periodic boundary conditions but no point periodicity. $\mathrm{ Mat_1}$ is assigned to the center inclusion, $\mathrm{ Mat_2}$ is assigned to sub-domain $\Omega_2$ and $ \mathrm{ Mat_3}$ is assigned to the matrix elsewhere}
    \label{fig4}
    \end{figure}

In the previous example, the geometry and properties were symmetrically distributed.
In this example, we consider a non-symmetric RVE.
The example consists of 9 different sub-domains as before.  However the material properties of the subdomain above the center inclusion are altered. Due to this, the symmetry of the problem is lost and 18 different amplitude vectors have to be considered. 
Fig.~\ref{fig4} shows the different sub-domains with the corresponding amplitude vectors.
    
The average total strains for the non-symmetric RVE are computed in a similar manner to the prior example and are given as
 \be
\begin{array}{l}
\label{eq54a}
\D \bfm{e}_1 = \bfm{e}_\mathrm{M} + \sym \left((\bfm{a}_6 - \bfm{a}_5) \otimes \bfm{n}_1 + (\bfm{a}_{15} - \bfm{a}_{14}) \otimes \bfm{n}_2\right), \\
\D \bfm{e}_2 = \bfm{e}_\mathrm{M} + \sym \left((\bfm{a}_9 - \bfm{a}_8) \otimes \bfm{n}_1 + (\bfm{a}_{13} - \bfm{a}_{15}) \otimes \bfm{n}_2\right), \\
\D \bfm{e}_3 =\bfm{e}_\mathrm{M} + \sym \left((\bfm{a}_2 - \bfm{a}_1) \otimes \bfm{n}_1 + (\bfm{a}_{11} - \bfm{a}_{10}) \otimes \bfm{n}_2\right), \\
\D \bfm{e}_4 = \bfm{e}_\mathrm{M} + \sym \left((\bfm{a}_3 - \bfm{a}_2) \otimes \bfm{n}_1 + (\bfm{a}_{14} - \bfm{a}_{13}) \otimes \bfm{n}_2\right), \\
\D \bfm{e}_5 = \bfm{e}_\mathrm{M} + \sym \left((\bfm{a}_1 - \bfm{a}_3) \otimes \bfm{n}_1 + (\bfm{a}_{17} - \bfm{a}_{16}) \otimes \bfm{n}_2\right), \\
\D \bfm{e}_6 =\bfm{e}_\mathrm{M} + \sym \left((\bfm{a}_5 - \bfm{a}_4) \otimes \bfm{n}_1 + (\bfm{a}_{12} - \bfm{a}_{11}) \otimes \bfm{n}_2\right), \\
\D \bfm{e}_7 = \bfm{e}_\mathrm{M} + \sym \left((\bfm{a}_4 - \bfm{a}_6) \otimes \bfm{n}_1 + (\bfm{a}_{18} - \bfm{a}_{17}) \otimes \bfm{n}_2\right), \\
\D \bfm{e}_8 = \bfm{e}_\mathrm{M} + \sym \left((\bfm{a}_8 - \bfm{a}_7) \otimes \bfm{n}_1 + (\bfm{a}_{10} - \bfm{a}_{12}) \otimes \bfm{n}_2\right), \\
\D \bfm{e}_9 = \bfm{e}_\mathrm{M} + \sym \left((\bfm{a}_7 - \bfm{a}_9) \otimes \bfm{n}_1 + (\bfm{a}_{16} - \bfm{a}_{18}) \otimes \bfm{n}_2\right).
\end{array}
\ee

We consider three different isotropic  materials within the composite. Mat$_\mathrm{1}$ in sub-domain {$\Omega_1$},  Mat$_\mathrm{2}$ in sub-domain {$\Omega_2$}, and Mat$_\mathrm{3}$ elsewhere in the matrix. The material parameters are given as follows: (Mat$_\mathrm{1}$) $E_1 = 1000 \,\mathrm{N/mm^{2}}$, $\nu_1=0.25$, $\sigma_{\mathrm{y}1} = 1.5 \,\mathrm{N/mm^{2}}$;  (Mat$_\mathrm{2}$) $E_2 = 2500 \,\mathrm{N/mm^{2}}$, $\nu_2=0.3$, $\sigma_{\mathrm{y}2} = 2.0 \,\mathrm{N/mm^{2}}$; and (Mat$_\mathrm{3}$) $E_3 = 5000 \,\mathrm{N/mm^{2}}$, $\nu_3=0.15$, $\sigma_{\mathrm{y}3} = 3.75 \,\mathrm{N/mm^{2}}$.   The loading protocol is the same as in the prior example.

\begin{figure}
\centering
  \includegraphics[scale=0.68]{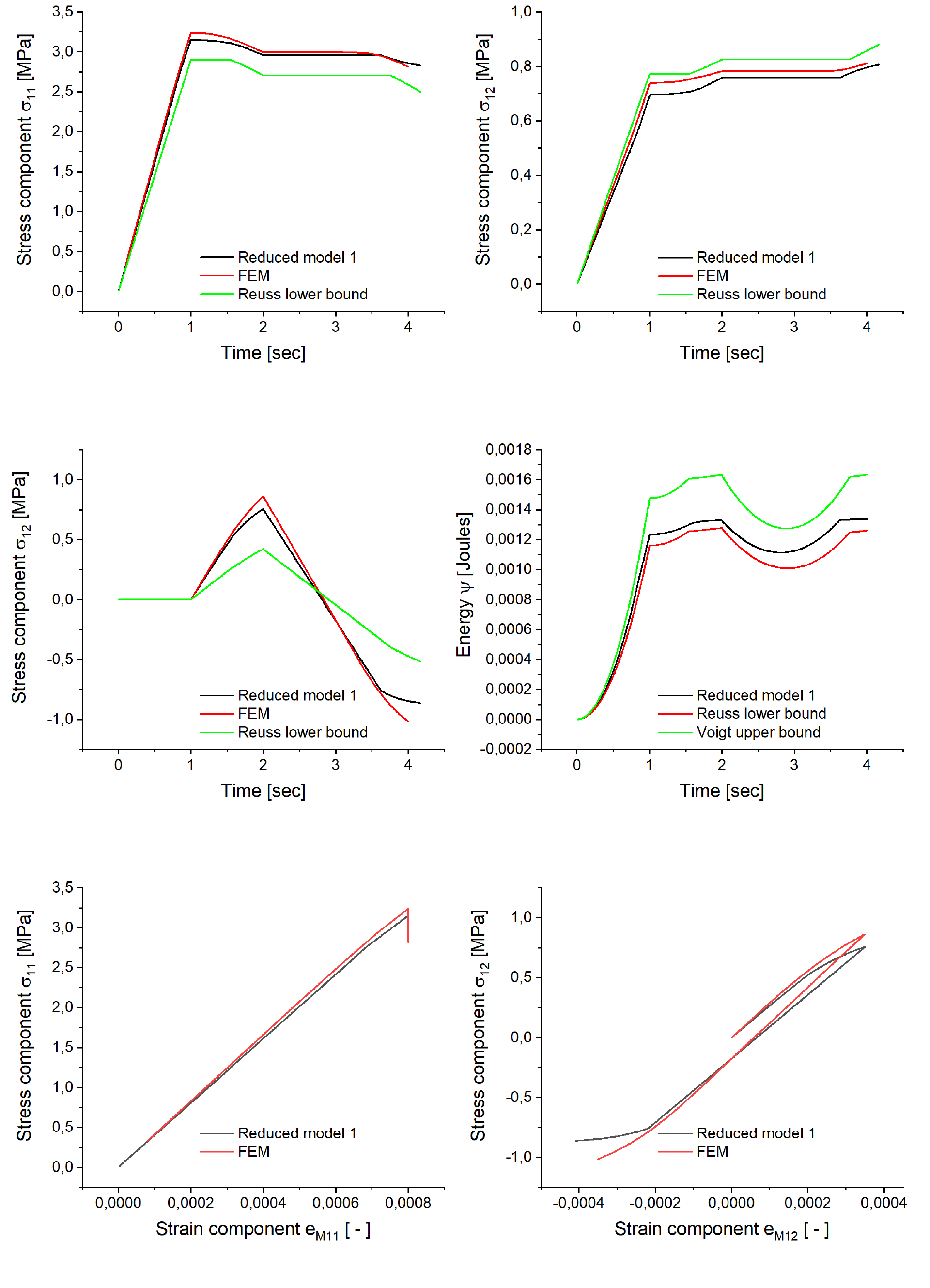}  
    \caption{Comparison of the effective model, a Reuss model, and a finite element model for a two inclusion RVE. 
    (top row and middle left pane)  macroscopic stresses versus time.  (middle right pane) RVE energy versus time.  (bottom row) stress-strain response for the normal stress and strain in the 1-direction and the shear stress and strain in the 1-2 plane.}
    \label{fig5}
      \end{figure}

Beginning with the macroscopic stress versus time response in Fig.~\ref{fig5}, it is seen that the reduced model provides a reasonably faithful macroscopic stress response for all in-plane components in mean stress space for the chosen loading path.  The Reuss lower bound model is seen to be reasonably accurate, though it somewhat fails to properly capture the shear behavior, as was true in the prior example.  Observing Fig.~\ref{fig5} (bottom row), it is also seen that the reduced model provides a surprisingly good prediction of the overall stress-strain response in this example in comparison to the much higher degree of freedom finite element model.

In Fig.~\ref{fig5} (middle right pane) the energy versus time curve for this example is plotted in comparison to the Voigt upper bound and Reuss lower bound. The location of the energy curve for the reduced model (red line) between the upper and lower bounds provides a good qualitative behavior that complies with Hill's condition. 

\begin{figure}
\centering
  \includegraphics[scale=0.62]{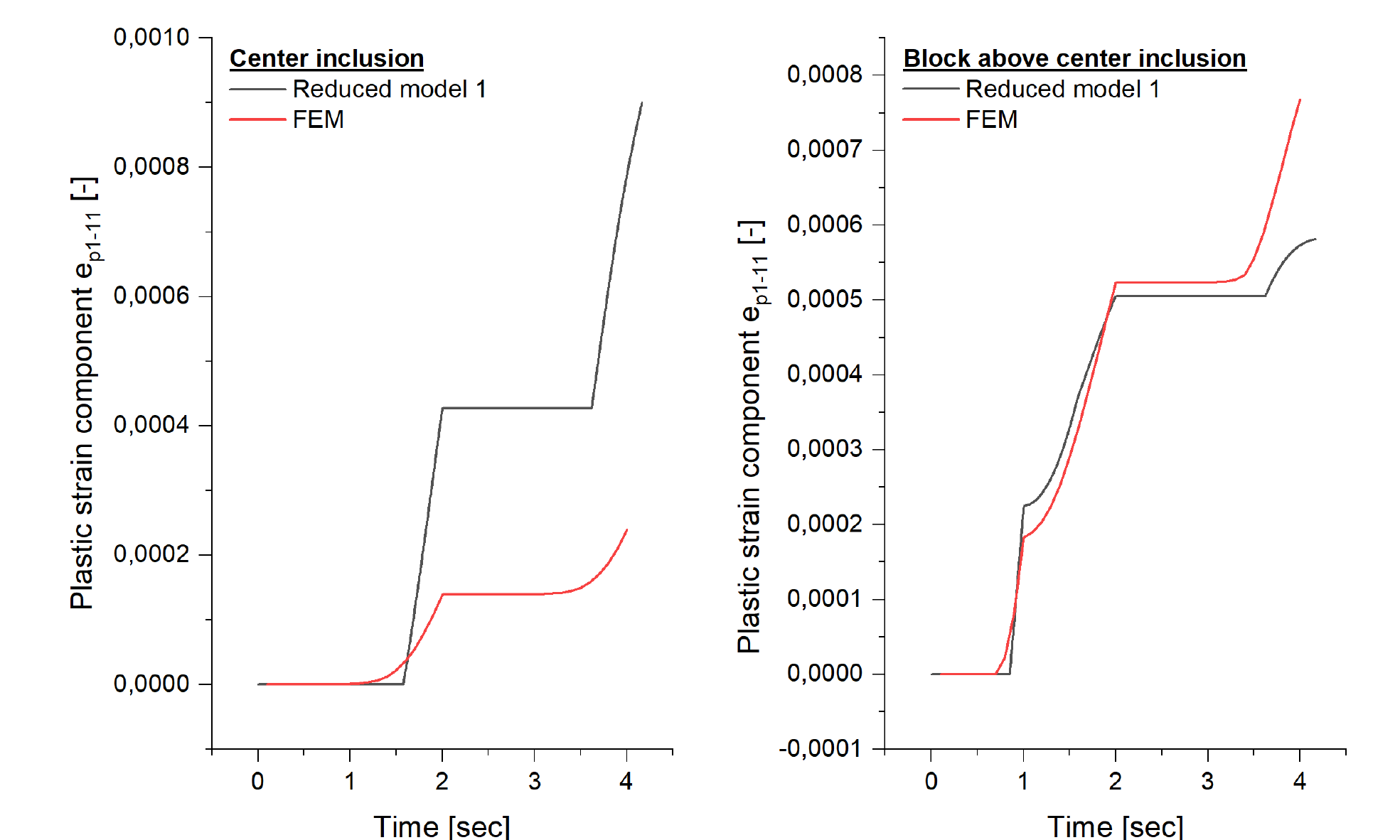}  
    \caption{Illustration to the plastic strain behavior averaged over $\mathrm{ Mat_1}$ at the center inclusion and $\mathrm{ Mat_2}$ at sub-domain $ \Omega_2$ in comparison to the finite element results.}
    \label{fig6}
      \end{figure}

In Fig.~\ref{fig6} we examine the 11 component of the mean plastic strain in the center inclusion and the sub-domain just above it.   As was seen in the prior example,  the response in the center inclusion is poorly represented by the reduced order model in comparison to the FEA computation.  However, at least for the 11 component, the reduced order model is able to capture the mean plastic behavior in $\Omega_2$.  The exact degree of capture of the local plastic behavior in the reduced model is an open question and is seen by just these two simple plots to be quite complex.  Notwithstanding, we still see quite reasonable macroscopic response from the reduced order model.

\newpage
\subsection{Two scale homogenization scheme including the reduced model}

\begin{figure}
\centering
  \includegraphics[scale=0.52]{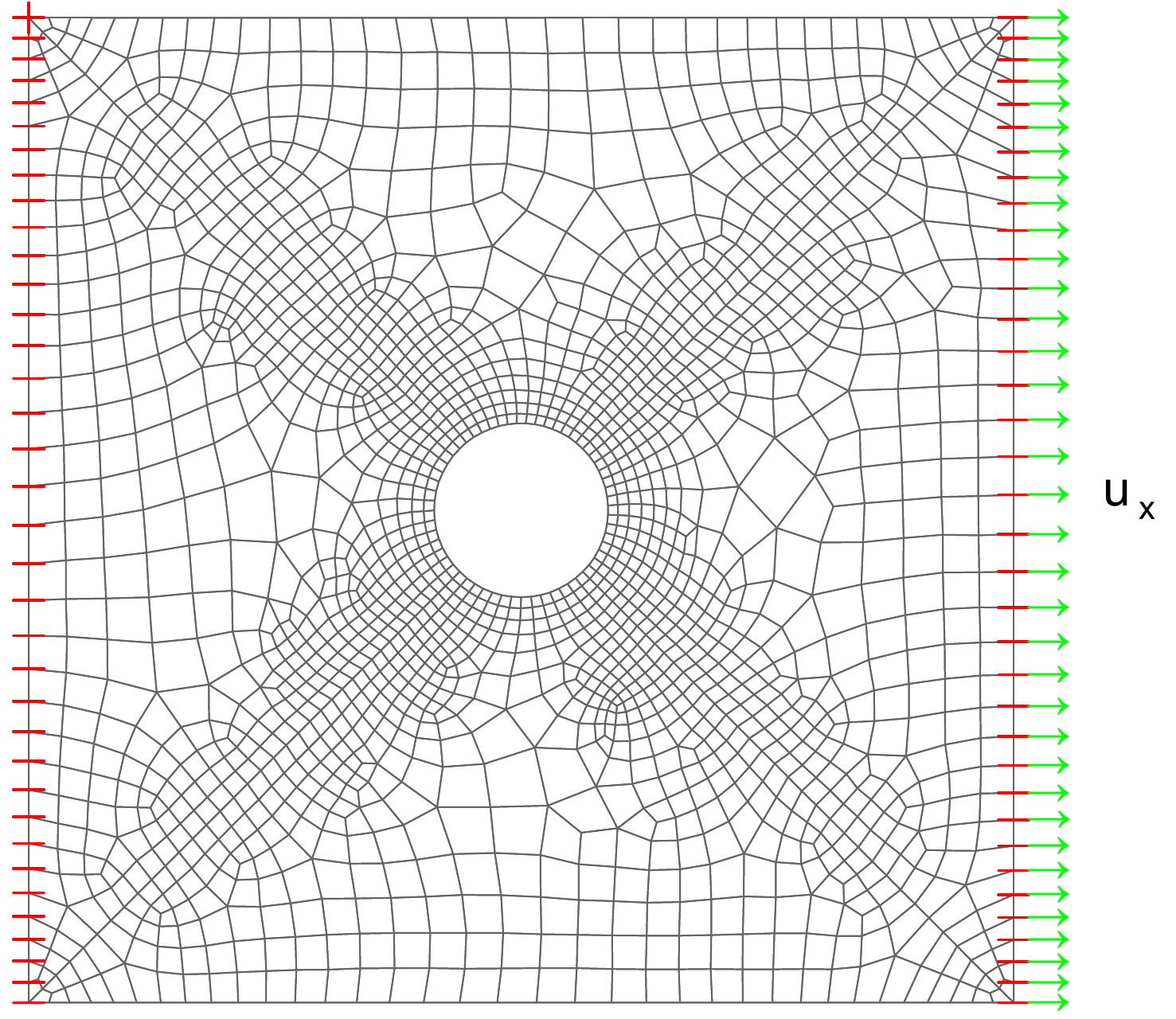}  
    \caption{A plate with a hole under given displacement at the right hand side.}
    \label{fig11a}
      \end{figure}

We now consider a well known boundary value problem, namely the plate with a hole. See Fig.~\ref{fig11a}.
We wish to consider a plate where the material has a well defined periodic microstructure. One well-understood computational framework for approaching such a problem is the two-scale FE$^2$ homogenization scheme.  In this scheme a finite element discretization is made of the macroscopic plate and then the pointwise stress strain response is computed using a finite element discretization of the microstructure, along with the Hill-Mandel procedure \citep[see e.g.][Chapter 7]{zt2n7}.  The scheme is generally considered to be computationally expensive in practice.  As an alternative, we consider the same setup, but one in which the microscale FE$^2$ response is replaced with an effective model as outline in the present paper.

\begin{figure}
\centering
  \includegraphics[scale=0.40]{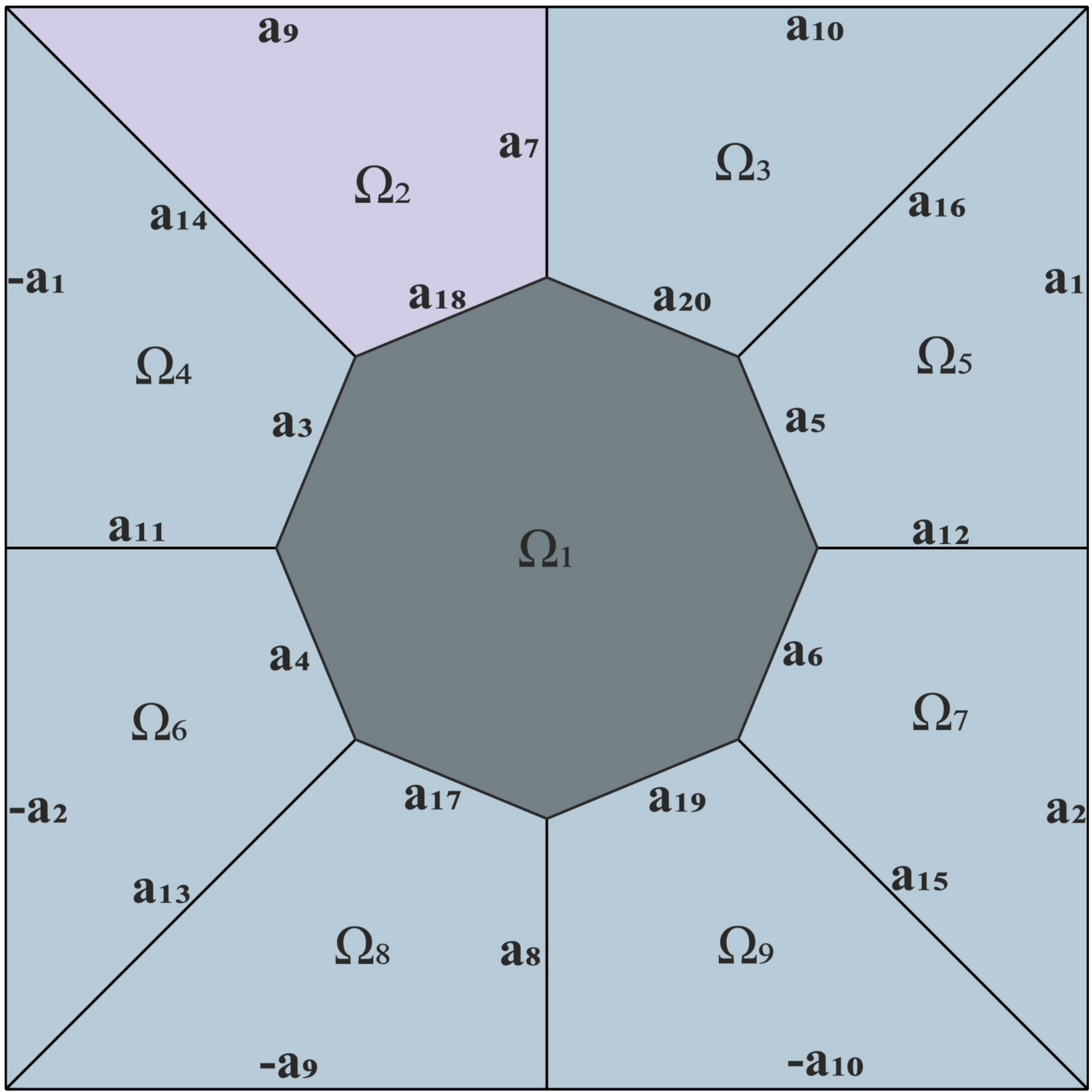}  
    \caption{Representative volume element with octagonal center inclusion and 20 amplitude vectors describing the different facets. $\mathrm{ Mat_1}$ is assigned to the center inclusion, $\mathrm{ Mat_2}$ is assigned to sub-domain $\Omega_2$ and $\mathrm{ Mat_3}$ is assigned elsewhere.}
    \label{fig12}
      \end{figure}

\begin{figure}
  \includegraphics[scale=0.55]{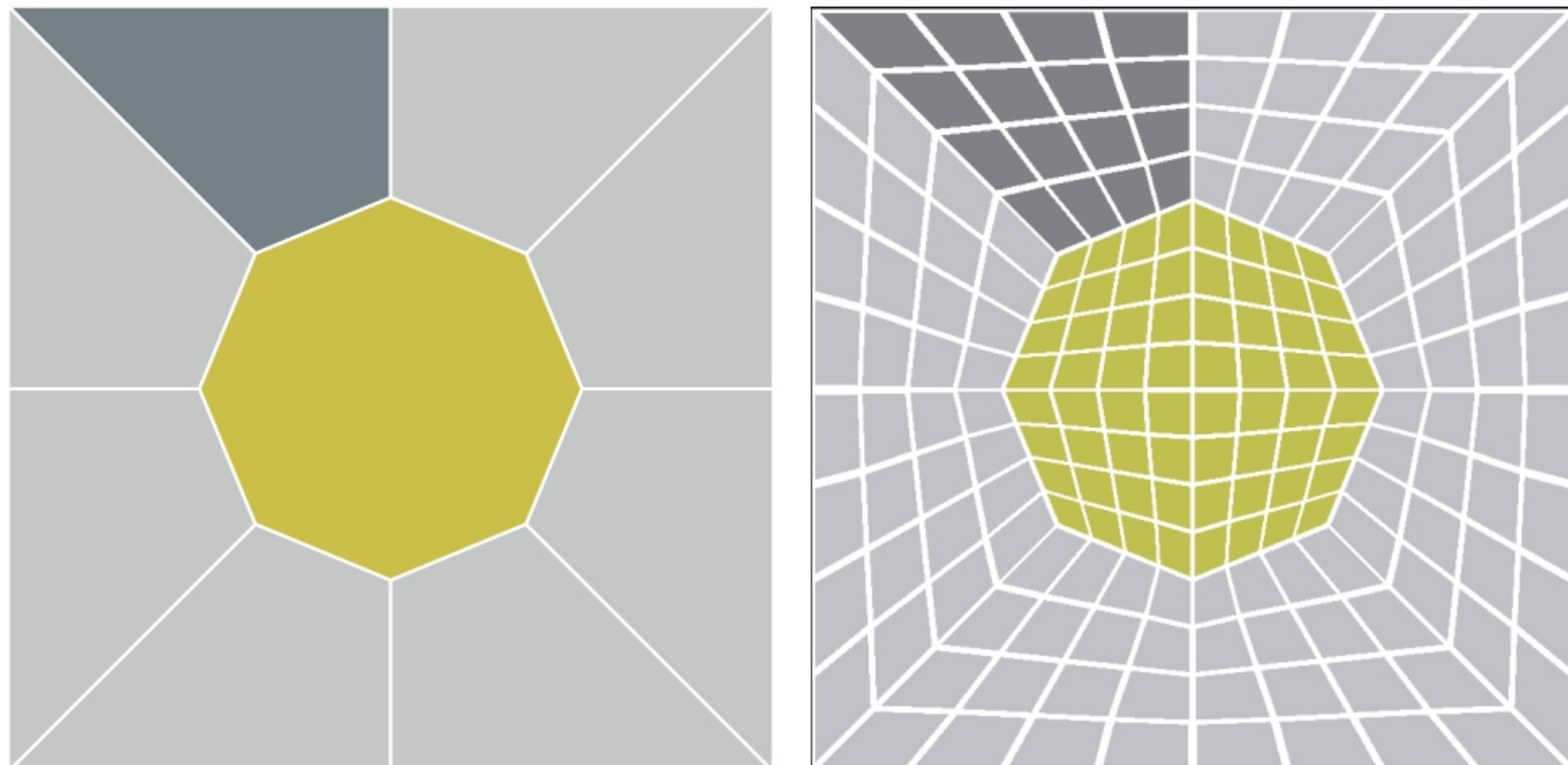}  
    \caption{RVE at the microstructure, with 9 sub-domains for the reduced model (left) and with 192 finite element mesh for the full model (right).}
    \label{fig16}
      \end{figure}

The microstructure is represented by the RVE shown in Fig.~\ref{fig12}.  For the effective model, there are 9 sub-domains and 20 amplitude vectors.  On the other hand, for the computation of the full FE$^2$ solution, the RVE is discretized into 192 tri-linear hexahedral elements with a fixed thickness implying the plane strain representation in FE$^2$FEAP \citep{feapfe2:20}; see Fig.~\ref{fig16}. Note that the computation results are scale invariant with respect to the microstructure.

The RVE itself consists of 3 different elastoplastic materials with \emph{linear hardening} -- Mat$_\mathrm{1}$ in sub-domain {$\Omega_1$}, Mat$_\mathrm{2}$ in sub-domain {$\Omega_2$},  and Mat$_\mathrm{3}$ elsewhere. The material parameters are given as follows:  (Mat$_\mathrm{1}$)  $E_1 = 1000 \,\mathrm{N/mm^{2}}$, $\nu_1=0.25$, $\sigma_{\mathrm{y}1} = 1.5 \,\mathrm{N/mm^{2}}$, $a_1=15\, \mathrm{N/mm^{2}}$; (Mat$_\mathrm{2}$) $E_2 = 2500 \,\mathrm{N/mm^{2}}$, $\nu_2=0.3$, $\sigma_{\mathrm{y}2} = 2.0 \,\mathrm{N/mm^{2}}$,  $a_2=25\, \mathrm{N/mm^{2}}$; and Mat$_\mathrm{3}$) $E_3 = 5000 \,\mathrm{N/mm^{2}}$, $\nu_3=0.15$, $\sigma_{\mathrm{y}3} = 3.75 \,\mathrm{N/mm^{2}}$, $a_3=30 \,\mathrm{N/mm^{(2)}}$.  For an explicit assessment of the octagonal inclusion RVE model, please see \cite{GGH:2021}.

The plate itself has dimensions 10 mm x 10 mm and possesses a centrally located hole of radius 1.25 mm. The left side of the plate is restrained from moving horizontally and the right side is subjected to a linearly increasing horizontal displacement, $U_x$, to a magnitude of 0.01 mm. At the macroscopic level the plate is discretized with 1793 tri-linear hexahedral elements restrained in $z-$direction, giving a plane-strain state.

\begin{figure}
\centering
  \includegraphics[scale=0.5]{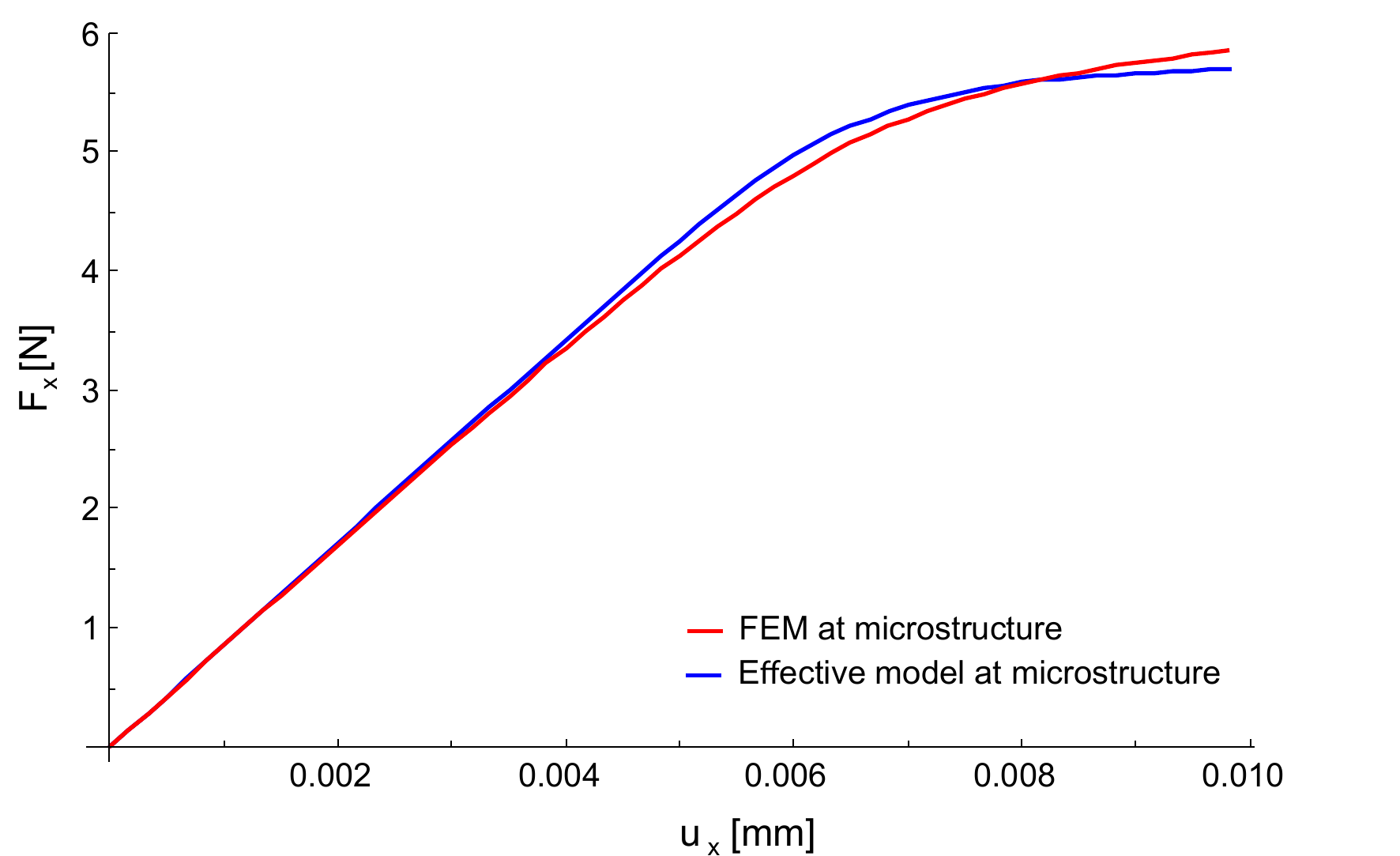}
    \caption{Force-Displacement diagram in $x$-direction obtained from the reduced form simulation and FE$^2$ method.}
    \label{fig17}
      \end{figure}

The horizontal force-displacement diagram computed from the response of the right side of the plate is shown in Fig.~\ref{fig17}, in which the behavior from the reduced simulation is well represented in comparison to the much higher degree of freedom FE$^2$ computations. We observe that the elastic compliance is well represented as is the basic yielding and plastic flow at the macroscopic level.
The matching compliances are a bit fortuitous.  Refinement of the FEA computation and/or using higher order elements would likely increase the FEA model's compliance, but clearly at an added computational cost.

An important advantage worth mentioning is the relative computation times for the two models. 
The computational time needed to complete the FE$^2$ was 71700\,sec, whereas on the same hardware it was only 8548\,sec -- a speed up of roughly 8.4 times.  This speed-up ratio could likely be improved via an optimized implementation of the effective model.  Notwithstanding, the resulting efficiency is clear.

\section{Conclusion}

We have presented a general variational methodology for the creation of effective or reduced models.
The method applies to structural reductions as well as reductions to describe composite bodies.  The central requirement is that the system's material constituents be described
via a free energy function and a dissipation potential.
The resulting reduced models possess a vastly reduced number of degrees of freedom, but are able to capture the essential features of the original systems. In contrast to typical order reduction schemes, the proposed method preserves the structure of the original system.  In particular, our reduced models are described by a coarse grained free energy and a coarse grained dissipation function, analogous to the fine scale material's description. Further, the internal evolutions are governed by a coarse grained Biot principle.

We have seen that for structural order reduction, the torsion and sphere examples, the method is quite effective and provides very good structural scale response for both perfectly plastic and linearly hardening elasoplastic materials.  The scheme also permits recovery of fine scale fields and these are seen to be reasonable though far from perfect.

Our methodology also proves its applicability to microstructural homogenization of elastic-perfectly plastic composites with periodic and non-periodic RVEs.  For this class of problems,
we found good agreement concerning macroscopic stress-strain response in comparison to a detailed finite element computation. Unsurprisingly, we saw larger deviations when examining the predicted microscale plastic strains, just as with the structural reduction problems. 
In our opinion, this behavior was mainly caused by a strong concentration of the plastic strains in small regions in the original models, that could not be matched by the effective ones.
Due to the coarseness, the effective models also struggle to match composite elastic stiffnesses.
Another drawback of the proposed method is, that it is applicable only to RVEs with polygonal sub-domains, as rounded or elliptical inclusions will require the use of an infinite number of boundary normals in the procedure we have outlined.
The use of higher order moments/variables in the definitions of the essential variables is one possible avenue of future research to resolve these last two points.

\section*{Acknowledgments}
SG's work was partially financed by the generous support of the Alexander von Humboldt Foundation.
KH and GJ gratefully acknowledge the funding by the German Research
Foundation (DFG) within the Priority Program 2256 Variational Methods
for Predicting Complex Phenomena in Engineering Structures and
Materials within project 441211072/441468770 Variational modeling of
pressure-dependent plasticity - a paradigm for model reduction via relaxation.

\section*{Bibliography}
\bibliography{klaus}

@article{GGH:2021,
	author = {Jezdan, Ghina and Govindjee, Sanjay and Hackl, Klaus},
	doi = {https://doi.org/10.1002/pamm.202100053},
	eprint = {https://onlinelibrary.wiley.com/doi/pdf/10.1002/pamm.202100053},
	journal = {PAMM},
	number = {1},
	pages = {e202100053},
	title = {Variational based effective models for inelastic materials},
	url = {https://onlinelibrary.wiley.com/doi/abs/10.1002/pamm.202100053},
	volume = {21},
	year = {2021}}

@manual{feapfe2:20,
	author = {R. L. Taylor and S. Govindjee},
	note = {\url{http://projects.ce.berkeley.edu/feap}},
	organization = {University of California, Berkeley.},
	title = {FEAP - {A} {Finite} {Element} {Analysis} {Program}, {FE$^2$} {Multiscale} {User} {Manual}},
	year = 2020}

@book{zt2n7,
	address = {Oxford},
	author = {Zienkiewicz, O. C. and Taylor, R. L. and Fox, D. D.},
	edition = {$7^{th}$},
	publisher = {Elsevier},
	title = {The Finite Element Method for Solid and Structural Mechanics},
	year = {2013}}

@manual{feapu:20,
	author = {R. L. Taylor and S. Govindjee},
	note = {\url{http://projects.ce.berkeley.edu/feap}},
	organization = {University of California, Berkeley.},
	title = {FEAP - {A} {Finite} {Element} {Analysis} {Program}, {User} {Manual}},
	year = 2020}

@book{rockafellar1970,
	author = {Rockafellar, R.T.},
	publisher = {Princeton University Press},
	series = {Princeton Landmarks in Mathematics and Physics},
	title = {Convex Analysis},
	year = {1970}}

@article{miehe.ea:04,
	author = {Miehe, C. and G\"oktepe, S. and Lulei, F.},
	journal = {Journal of the Mechanics and Physics of Solids},
	pages = {2617-2660},
	title = {A micro-macro approach to rubber-like materials -- {P}art {I}: The non-affine micro-sphere model of rubber elasticity},
	volume = 52,
	year = 2004}

@book{govindjee2013,
	author = {Govindjee, S.},
	location = {Cambridge},
	publisher = {Oxford University Press},
	title = {Engineering Mechanics of Deformable Solids: {A} Presentation with Exercises},
	year = 2013}

@article{govindjee2007upper,
	author = {Govindjee, S. and Hackl, Klaus and Heinen, Rainer},
	doi = {10.1007/s00161-006-0038-1},
	issn = {0935-1175},
	journal = {Continuum Mech. Therm.},
	number = {7-8},
	pages = {443-453},
	title = {An upper bound to the free energy of mixing by twin-compatible lamination for n-variant martensitic phase transformations},
	volume = {18},
	year = {2007}}

@article{hackl2008micromechanical,
	author = {Hackl, Klaus and Heinen, Rainer},
	doi = {10.1007/s00161-008-0067-z},
	issn = {0935-1175},
	journal = {Continuum Mech. Therm.},
	pages = {8},
	title = {A micromechanical model for pretextured polycrystalline shape-memory alloys including elastic anisotropy},
	volume = {19},
	year = {2008}}

@article{hackl2012zamm,
	author = {Hackl, K. and Heinz, S. and Mielke, A.},
	doi = {10.1002/zamm.201100155},
	issn = {1521-4001},
	journal = {ZAMM},
	number = {11-12},
	pages = {888--909},
	publisher = {WILEY-VCH Verlag},
	title = {A model for the evolution of laminates in finite-strain elastoplasticity},
	url = {http://dx.doi.org/10.1002/zamm.201100155},
	volume = {92},
	year = {2012}}

@article{junker2011finite,
	author = {Junker, P. and Hackl, K.},
	journal = {Computational Mechanics},
	number = {5},
	pages = {505--517},
	publisher = {Springer-Verlag New York, Inc.},
	title = {Finite element simulations of poly-crystalline shape memory alloys based on a micromechanical model},
	volume = {47},
	year = {2011}}

@article{kochmann2011evolution,
	author = {Kochmann, D.M. and Hackl, K.},
	journal = {Continuum Mech. Therm.},
	number = {1},
	pages = {63--85},
	publisher = {Springer},
	title = {The evolution of laminates in finite crystal plasticity: a variational approach},
	volume = {23},
	year = {2011}}

@article{miehe2003,
	author = {Miehe, Christian and Lambrecht, Matthias},
	doi = {10.1002/nme.726},
	issn = {1097-0207},
	journal = {Int. J. Numer. Meth. Engrg.},
	number = {1},
	pages = {1--41},
	publisher = {John Wiley & Sons, Ltd.},
	title = {Analysis of microstructure development in shearbands by energy relaxation of incremental stress potentials: Large-strain theory for standard dissipative solids},
	url = {http://dx.doi.org/10.1002/nme.726},
	volume = {58},
	year = {2003}}

@article{miehe2004,
	author = {C. Miehe and M. Lambrecht and E. G{\"u}rses},
	doi = {10.1016/j.jmps.2004.05.011},
	issn = {0022-5096},
	journal = {J. Mech. Phys. Solids},
	number = {12},
	pages = {2725 - 2769},
	title = {Analysis of material instabilities in inelastic solids by incremental energy minimization and relaxation methods: evolving deformation microstructures in finite plasticity},
	url = {http://www.sciencedirect.com/science/article/pii/S002250960400105X},
	volume = {52},
	year = {2004}}

@article{miehe2010,
	author = {Frankenreiter, Ilona and Rosato, Daniele and Miehe, Christian},
	doi = {10.1002/pamm.201010138},
	issn = {1617-7061},
	journal = {PAMM},
	number = {1},
	pages = {291--292},
	publisher = {WILEY-VCH Verlag},
	title = {A Hybrid MicroMacroModel for the Description of Evolving Anisotropy in Finite Polycrystal Plasticity},
	url = {http://dx.doi.org/10.1002/pamm.201010138},
	volume = {10},
	year = {2010}}

@article{miehe2011a,
	author = {E. G{\"u}rses and C. Miehe},
	doi = {10.1016/j.jmps.2011.01.002},
	issn = {0022-5096},
	journal = {J. Mech. Phys. Solids},
	number = {6},
	pages = {1268 - 1290},
	title = {On evolving deformation microstructures in non-convex partially damaged solids},
	url = {http://www.sciencedirect.com/science/article/pii/S0022509611000032},
	volume = {59},
	year = {2011}}

@article{ContiDolzmannKlust2009,
	author = {S. Conti and G. Dolzmann and C. Klust},
	journal = {Proc. Roy. Soc. A},
	link = {http://dx.doi.org/10.1098/rspa.2008.0390},
	pages = {1735-1742},
	title = {Relaxation of a class of variational models in crystal plasticity},
	volume = 465,
	year = {2009}}

@article{ContiOrtiz05,
	author = {S. Conti and M. Ortiz},
	journal = {Arch. Rat. Mech. Anal.},
	link = {http://www.springerlink.com/openurl.asp?genre=article&id=doi:10.1007/s00205-004-0353-2},
	pages = {103-147},
	title = {Dislocation Microstructures and the Effective Behavior of Single Crystals},
	volume = 176,
	year = {2005}}

@article{Aubry20032823,
	author = {Sylvie Aubry and Matt Fago and Michael Ortiz},
	doi = {10.1016/S0045-7825(03)00260-3},
	issn = {0045-7825},
	journal = {Comput. Methods Appl. Mech. Engrg.},
	number = {26-27},
	pages = {2823 - 2843},
	title = {A constrained sequential-lamination algorithm for the simulation of sub-grid microstructure in martensitic materials},
	url = {http://www.sciencedirect.com/science/article/pii/S0045782503002603},
	volume = {192},
	year = {2003}}

@article{Dvorak1992,
	author = {George J. Dvorak and Yakov Benveniste},
	journal = {The Royal sociaty},
	link = {http://rspa.royalsocietypublishing.org/},
	pages = {311-327},
	title = {On transformation strains and uniform fields in multiphase elastic media},
	volume = 437,
	year = {1992}}

@article{DvorakBenveniste1992,
	author = {George J. Dvorak},
	journal = {The Royal sociaty},
	link = {http://rspa.royalsocietypublishing.org/},
	pages = {291-310},
	title = {Transformation field analysis of inelastic composite materials},
	volume = 437,
	year = {1992}}

@article{MichelSuquet2003,
	author = {J.C. Michel and P. Suquet},
	doi = {10.1016/S0020-7683(03)00346-9},
	journal = {International journal of solids and structures},
	pages = {6937-6955},
	title = {Nonuniform transformation field analysis},
	volume = 40,
	year = {2003}}

@article{WAIMANN2016110,
	author = {Johanna Waimann and Philipp Junker and Klaus Hackl},
	doi = {https://doi.org/10.1016/j.euromechsol.2015.08.001},
	issn = {0997-7538},
	journal = {European Journal of Mechanics - A/Solids},
	pages = {110 - 121},
	title = {A coupled dissipation functional for modeling the functional fatigue in polycrystalline shape memory alloys},
	url = {http://www.sciencedirect.com/science/article/pii/S0997753815001011},
	volume = {55},
	year = {2016}}

@article{gunther2015variational,
	author = {G{\"u}nther, Christina and Junker, Philipp and Hackl, Klaus},
	journal = {Proceedings of the Royal Society A: Mathematical, Physical and Engineering Sciences},
	number = {2180},
	pages = {20150110},
	publisher = {The Royal Society Publishing},
	title = {A variational viscosity-limit approach to the evolution of microstructures in finite crystal plasticity},
	volume = {471},
	year = {2015}}

@article{junker2014thermo,
	author = {Junker, Philipp and Hackl, Klaus},
	journal = {Continuum Mechanics and Thermodynamics},
	number = {6},
	pages = {859--877},
	publisher = {Springer Berlin Heidelberg},
	title = {A thermo-mechanically coupled field model for shape memory alloys},
	volume = {26},
	year = {2014}}

@article{govindjee2019fully,
	author = {Govindjee, Sanjay and Zoller, Miklos J and Hackl, Klaus},
	journal = {Journal of the Mechanics and Physics of Solids},
	pages = {1--19},
	publisher = {Pergamon},
	title = {A fully-relaxed variationally-consistent framework for inelastic micro-sphere models: Finite viscoelasticity},
	volume = {127},
	year = {2019}}

@article{FRITZEN2015114,
	author = {Felix Fritzen and Sonia Marfia and Valentina Sepe},
	doi = {https://doi.org/10.1016/j.compstruc.2015.05.012},
	issn = {0045-7949},
	journal = {Computers \& Structures},
	pages = {114 - 131},
	title = {Reduced order modeling in nonlinear homogenization: A comparative study},
	url = {http://www.sciencedirect.com/science/article/pii/S0045794915001492},
	volume = {157},
	year = {2015}}





\end{document}